\documentclass{article}
\usepackage{graphicx}
\usepackage{subfig}
\usepackage{float}
\usepackage{amssymb,amstext,amsmath,amsfonts}
\usepackage{fullpage}
\usepackage{color}

\usepackage{cite}

\allowdisplaybreaks
\makeatletter

\renewcommand{\@maketitle}{
\newpage
\null
\vskip 2em%
\begin{center}%
{\LARGE \@title \par}%
\end{center}%
\begin{center}%
{\large \@author \par}%
\vskip 1em%
{\large Los Alamos National Laboratory, MS F644, Los Alamos, NM 87545 \par}%
\end{center}%
\par} \makeatother

\begin{document}

\title{Plasma Viscosity with Mass Transport in Spherical ICF Implosion Simulations}

\author { E. L. Vold$^1$, A. S. Joglekar$^2$, M. I. Ortega$^3$, R. Moll$^4$, D. Fenn$^5$, K. Molvig$^1$\\
$^1$Los Alamos National Laboratory, Los Alamos, NM, USA\\
$^2$U. of Michigan, Ann Arbor, MI, USA\\
$^3$U. of New Mexico, Albuquerque, NM\\
$^4$U. of California, Santa Cruz, CA, USA\\
$^5$Florida State U., Tallahassee, FL, USA\\
}

\date{\today}

\maketitle

\begin{abstract}
\noindent The effects of   viscosity and small-scale atomic-level mixing on plasmas in inertial confinement fusion (ICF) currently represent challenges in ICF research. Many current ICF hydrodynamic codes ignore the effects of viscosity though recent research indicates viscosity and mixing by classical transport processes may have a substantial impact on implosion dynamics. We have implemented a Lagrange hydrodynamic code in one-dimensional spherical geometry with plasma viscosity and mass transport  and including a three temperature model for ions, electrons, and radiation treated  in a gray radiation diffusion approximation.  The code is used to study ICF implosion differences with and without  plasma viscosity and to determine the impacts of viscosity on temperature histories and neutron yield. It was found that plasma viscosity has substantial impacts on ICF shock dynamics characterized by shock burn timing, maximum burn temperatures, convergence ratio, and time history of  neutron production rates.   Plasma viscosity reduces the need for artificial viscosity to maintain numerical stability in the Lagrange formulation and also modifies the flux-limiting needed for electron thermal conduction. 
\end{abstract}

\section{Introduction}

Direct-drive inertial confinement fusion (ICF) refers to laser heating of a  spherical shell that contains fusion fuel, resulting in compression and fusion \cite{Atzenitext, soures1996direct}. Perfect spherical compression has many shortcomings resulting from drive asymmetries, laser-plasma  instabilities, and fluid  instabilities including Rayleigh-Taylor (RT), Richtmyer-Meshkov (RM) and Kelvin-Helmholtz (KH). Target capsule non-uniformities  as well as beam-to-beam imbalances become exaggerated during implosion convergence and perfect spherical compression is compromised and degrades performance. The implosion dynamics can also be affected by the shell material mixing into the fuel, and  the mix can in turn be influenced by plasma viscosity.
Plasma effects on burn physics and implosions have been the focus of many studies, including work on the plasma mix layer structure \cite{molvigvolddoddwilks}, barodiffusion \cite{amendt2011potential}, flux limiters \cite{schurtz, schurtz2007revisiting, marocchino2014effects}, as well as laser absorption mechanisms \cite{radha2005multidimensional} and asymmetries \cite{Theobald} mostly using hydrodynamics codes in comparison to experimental data.  

There has been increased recent interest in plasma kinetics as a possible mechanism to explain aspects of ICF implosion degradation \cite{rinderknecht, HoffmanEtAl}. 
Kinetic theory describes plasma transport in the small Knudsen number limit resulting from particle collisions and attempts to  determine consistent  transport coefficients multiplying the known gradient quantities to determine species diffusivity, viscosity and thermal conduction for the mass, momentum, and energy conservation equations \cite{gas metal, Simakov, hinton}. 
In ICF simulations, energy transport typically includes the classical species temperature coupling and  thermal conduction  [i.e., Spitzer \cite{spitzer}].  Conduction is dominated by the electron contribution and is often flux limited as a mechanism to improve simulation results in comparison to experimental data  \cite{schurtz2007revisiting, marocchino2014effects}.  Species plasma  viscosity is usually assumed to be negligible \cite{Atzeni87, Atzenitext} and may be overwhelmed by artificial viscosity  \cite{wilkins} used in many Lagrangian-based hydrodynamic  simulation codes  for numerical stability and to capture the shock discontinuities. Particle diffusion is often assumed to be negligible while species mixing is modeled as  the result of classical fluid instabilities (RT, RM, KH) \cite{Atzenitext} or  from turbulent processes \cite{ThomasKares, hainesicf}.  Recent analysis \cite{WilsonEtAl} suggests that the mix in ICF implosions involves comparable contributions from  chunk mix (e.g., subgrid scale fluid structures) and atomic mix, suggesting that plasma transport,  in viscosity and in species diffusion, may have  significant contributions to mix.

While previous studies examine various physics issues that may affect implosion dynamics, few specifically address the effect of plasma ion viscosity on ICF implosions.  Early on, Yabe and Tanaka \cite{Yabe} reported that plasma viscosity is likely to be important in ICF implosions, and following up on that,  simulations confirmed that kinetic effects may increase  the viscosity at high Mach number shocks \cite{Vidal}.    In a more recent  study,  \cite{manheimer2007effects} post-processing of ICF simulations was used to infer that viscous stress is small compared to the plasma kinetic pressures.  In another study, post-processing 'Omega-scale' ICF simulations, \cite{vold-sherrill}, the time integrated  viscous diffusive lengths were  estimated to reach the dimensions of the compressed fuel  during the maximum compression.  The differences and the uncertainties in transport estimates made by post-processing Eulerian simulations highlight the sensitivity of the results to the dynamic trajectories of the plasma densities and temperatures while also implying a need for self-consistent in-line plasma transport calculations.   

More recently, the role of plasma viscosity in smoothing small scale  fluctuations  in 2D ICF simulations has been reported \cite{weber2014inhibition}, where a distinct smoothing of small scale structures is evident in the ICF fuel region when viscosity is included.  Plasma viscosity and diffusion have been shown in 2D simulations to play a significant role in attenuating RT and KH interfacial instabilities for mode scale lengths and conditions relevant  in ICF \cite{plasmaInRTnKH}.  These studies  suggest  that  plasma dissipative processes are important at micron scale lengths for ICF relevant conditions.  This in turn suggests that high resolution 3D ICF simulations \cite{ThomasKares, hainesicf}  that now solve the Euler equations on sub-micron scale grids should include the plasma dissipative processes in the simulations for accuracy at the smallest scales.

In a study examining  the physical plasma viscosity  \cite{mason2014real}  and comparing to Von Neumann artificial viscosity \cite{wilkins},  it was found that the plasma viscosity in 2D geometries and conditions relevant to ICF is likely to be important as viscous dissipation can increase the fuel entropy and degrade implosion performance.  Similarly, our present study  implements plasma viscosity in addition to artificial viscosity.  We found that the computed results remain numerically stable if artificial viscosity is turned off after the first shock converges on center.  The solutions, with  artificial viscosity set to zero after the first shock convergence, lead to greater  fluctuations in density and in temperatures.  Assuming these fluctuations reduce the net radial implosion velocities, this implies one might expect a decrease in predicted performance using Lagrange codes with no artificial viscosity.

In the present study, we explore the dynamic  effects of viscosity on ICF implosions through the use of a one dimensional, three temperature, Lagrangian hydrodynamics model that also includes a treatment of fuel-plastic mass mixing by plasma transport.  The study focuses on  'Omega-facility' scale ICF experiments with simplified capsules of a CH shell and a deuterium fuel region.  It is  shown that the simulation results are generally consistent with experiments and previous simulations.  We then examine  A-B comparisons  of simulations both with and without plasma viscosity in the hydrodynamic equations.  We  demonstrate a significant influence on timing in 1D implosions, particularly at first shock convergence and burn conditions, while   yield and time-integrated burn-weighted temperature show less sensitivity to viscosity.

\section{Theory}

\subsection{Hydrodynamics}

The plasma is modeled with  one-dimensional  hydrodynamics equations for the mixture averaged quantities,

\begin{align}
 \frac{D \rho}{D t} &= -\rho \nabla \cdot \textbf{u} \label{density-eq} \\
 \rho \frac{D \textbf{u}}{D t} &= -\nabla p + \nabla \cdot (\eta_o \nabla \textbf{u}) \label{momentum-eq} \\
 \frac{n}{\gamma-1}{ \frac{D T}{D t} } &= -p \nabla \cdot \textbf{u} + \nabla \cdot (\kappa_e \nabla T) + u:\nabla \cdot (\eta_o \nabla \textbf{u})
\label{energy-eq}
\end{align}
where $\gamma = 5/3$ for the plasma  as a monatomic ideal gas, and $n$ is the number density for particles including ions and electrons.  
Equations  (\ref{density-eq}) to  (\ref{energy-eq})  express the conservation  of density and momentum with the evolution of temperature, respectively in the Lagrange frame.  Note that the latter two equations have  diffusive-like terms arising from the plasma transport,  a viscous stress tensor in Eq. (\ref{momentum-eq}) and the heat conduction in Eq. (\ref{energy-eq}).  The last term in  Eq. (\ref{energy-eq}) represents the viscous heat dissipation.  Enthalpy carried by the  species mass flux  contributes to the energy within the narrow  fuel-plastic mix layer formed by mass diffusion, however, this heating term is ignored in this study focusing on the viscous effects.

In the fuel, the transport coefficients are appropriate to a simple singly charged ion plasma  \cite{Braginskii},   $\eta_o = 0.96 nkT\tau_i$ and $\kappa_e = 3.2 nkT\tau_e$ with the collision times as given  in \cite{Braginskii, nrl}.  The viscosity is modified in the  higher z capsule material and  in the binary mixing layer between fuel and the CH capsule, consistently with the low z - high z kinetic formulation in  \cite{gasmetal}.    The approximate viscosity in the mix layer, with subscript $1$ representing the fuel ions,  and $2$ the average ion in the CH,  is evaluated as

\begin{equation}
\eta_o \approx  \frac{n_i kT_i} {\nu_i}  \approx  \frac{n_i kT_i}  {\sum_{j} \nu_{i,j} } \approx  \frac{n_i kT_i}  {C_\nu ( \nu_{1,1} + \nu_{1,2}  )}  \approx  \frac{n_i kT_i^{5/2}}  {C_\nu ( n_{1} + n_{2} z_2^2  )}
\end{equation}
where $z_1 = 1$ and $n_i = n_1 + n_2$. As a further approximation, the coulomb logarithm is assumed to be fixed  for each pairwise collision rate, $L \approx 5$, and is folded into the coefficient, $C_\nu$.  With the collision rates proportional to $kT_i^{-3/2}$, the  plasma viscosity scales as $kT_i^{5/2}$, and the kinematic viscosity scales approximately as $kT_i^{5/2} / n_i$, highlighting the sensitivity of the dynamically evolving viscosity to the local and instantaneous plasma ion temperature and density.

A three temperature model described in a following section provides  a more detailed description of the plasma.  As such, Eqn.\ref{energy-eq} is modified to include separate temperatures for the three components in the plasma: ions, electrons, and radiation. Thermal conductivities for ions or electrons in the 3T model scale as $T^{5/2}$ similarly to the ion viscosity.

\subsection{Species Mass Mixing}

In addition to the single fluid hydrodynamics, the code also tracks species mass composition.
The equation for the  mass fraction, $Y$, for the light ion species, $1$,  is 

\begin{equation}
\frac{ D Y } { Dt} = \rho^{-1} \nabla \cdot \rho_1 w_1
\end{equation}
where $w_1 = u_1 - u$ is a species drift flux, the species average velocity relative to the mixture mass averaged velocity, $u$.  The plasma species mass drift flux in binary mixing due to the concentration gradient and barodiffusion terms is written for the light ion  species, $1$, as

\begin{equation}
 \rho_1 w_1 \equiv m_1 \Gamma_1 = - \frac{ p_{ia} \rho_2 } {\rho \nu_{12} \alpha^k_{11}[\chi] } 
 \left(  \nabla \chi + ( \chi - Y ) \frac{\nabla p_{ia}} {p_{ia}}    + (Z - Y ) \frac{\nabla p_e} {p_{ia}} +  
  \chi \alpha_{Te}^K  \frac{\nabla Te_1} {Ti}    + \chi ( 1 - \chi) \alpha_{Ti}^K \frac{\nabla Ti_1} {Ti}  \right)
 \label{massmix}
\end{equation}
where $Y$ is the mass fraction of the light species, $p_{ia}$, is the total ion pressure, $\rho_2$ is the mass density of the heavier species, $\nu_{12}$ is a momentum exchange rate between the two species.  $\chi$ is the number fraction of the lighter species, and the charge fraction is $Z = n_1 z_1 / n_e$.  The transport coefficient, 
$\mu \equiv ( p_{ia} \rho_2 ) / ( \rho \nu_{12} \alpha^k_{11} [\chi] )$ can be shown  to be equivalent to that derived in \cite{gasmetal} with the kinetic correction factor, $\alpha^k_{11} [\chi] $.  The molar concentration gradient is converted to a mass fraction gradient to allow an implicit solution to the diffusive component of mass flux, using $\nabla \chi = (D \chi / D Y ) \nabla Y$, where $D \chi / D Y $ is a function of only the particle masses. The form for the first three terms in Eqn. (\ref{massmix}), the molar gradient, the ion barodiffusion, and electron barodiffusion, are  in general  agreement in the literature  \cite{gasmetal, KaganTang1, HoffmanEtAl}.  

The  thermal gradient forces, the last two terms in Eq. (\ref{massmix}), have recently been defined in the  literature in varying forms  \cite{gasmetal}, \cite{KaganTang2} \cite{HoffmanEtAl}.  
The thermal force coefficients here, $\alpha_{Te}^K, \alpha_{Ti}^K$ can be evaluated in terms of the kinetic transport coefficients, $\alpha_{11}, \alpha_{T}, \alpha_{12}$, given in \cite{gasmetal}, which modify the  coefficients for the concentration varying across the mix layer.  Although the coefficients of the temperature gradient force can appear to be large \cite{KaganTang2} our preliminary analysis suggests that when realistic temperature gradients  are accounted for,  then the forces are small in our test problems.  

\subsection{Plasma Pressure}

The plasma pressure utilizes an ideal gas approximation $p = nkT$, appropriate to multi-species, as a sum over ion species assumed to be fully ionized ideal  gases and assuming   ambipolar electrons.  The pressure can thus be expressed as a function of the total density, temperature, partial ionization, $z_i$, and the light ion mass fraction, $Y$.  For a single temperature and binary mix of ions with light ion  atomic mass, $A_1$, and heavy ion atomic mass, $A_2$, the pressure is 

\begin{equation}
p = n k T = (n_i + n_e ) k T = n_1(1 + z_1) + n_2(1+z_2) kT = \left(  \frac{Y}{A_1} (1+z_1) + \frac{(1-Y)}{A_2} (1+z_2) \right) \rho kT
\end{equation}
 and for the 3T model with $T_e \ne T_i$, 
\begin{equation}
p = n_i kT_i + n_e k T_e =   \rho  \left(  \frac{Y}{A_1} (kT_i + z_1 kT_e) + \frac{(1-Y)}{A_2} (kT_i + z_2 kT_e) \right)
\end{equation}


\subsection{Three Temperature Model for Ions, Electrons, and Radiation}

 In the 3T model, the temperature in  Eq. (\ref{energy-eq}) is replaced by temperatures for ions, electrons, and radiation, represented by three coupled  differential equations, 

\begin{align}
\label{Ti_eqn}
\frac{n_{i}}{\gamma -1 } k \frac{dT_{i}}{dt} &= -(p_i + q + \eta_o \nabla \vec{u}) \nabla \cdot \vec{u}+ \nabla \cdot (\kappa_{i} \nabla T_{i}) + E_{p} R_{DD} f_{pi} + \omega_{ie} (T_{e}-T_{i}) \\
\frac{n_{e}}{\gamma -1 } k \frac{dT_{e}}{dt} &= -p_{e} \nabla \cdot \vec{u} + \nabla \cdot (\kappa_{e} \nabla T_{e}) + E_{p} R_{DD} f_{pe} + \omega_{ie} (T_{i}-T_{e}) + (\omega_{p} + \omega_{c}) (T_{r} - T_{e}) + S_{laser} \\
\frac{16 \sigma_{SB} T_{r}^{3}}{c} \frac{dT_{r}}{dt} &= - p_{r} \nabla \cdot \vec{u} + \nabla \cdot (\kappa_{r} \nabla T_{r}) + (\omega_{p} + \omega_{c}) (T_{e} -T_{r}) \label{tempwork}
\end{align}\\
where $T_{i}, T_{e}, $ and $T_{r}$ are the ion, electron, and radiation temperatures respectively. Heat sources  include energy deposition to the electrons due to the laser heating, $S_{laser}$, nuclear fusion particle energy deposition $E_{p}$ (negligible in these simulations), heat conduction $k \nabla T $, adiabatic compression heating and non-adiabatic compression heating of the ions due to artificial viscosity, q, and the viscous heating, proportional to $\eta_o \nabla \vec{u}$.  The electron conduction is not flux limited but the heat flux by classical conduction will be  compared to the flux limited values as discussed  in the Results.

The deuterium-deteurium nuclear fusion reactions evolve by

\begin{equation}\label{1}
\frac{d n_{D}}{dt} = -\frac{1}{4}n_{D}^{2}<\sigma v>_{DD} 
\end{equation}
where $n_{D}$ is the deuterium number density.

\textit{Radiation transport coefficient: Rossland Opacity.}  
The radiative diffusion coefficient, $\kappa_r$, in Eq. (\ref{tempwork}) is based on an approximation to the grey-body diffusive flux, $\Gamma_r$, for the radiation energy density, $E_r = ( 4 \sigma_{SB} / c ) T_r^4$, as

\begin{equation}
\Gamma_r \approx - \frac{c \lambda_R} {3} \nabla E_r =  - \frac{4 \lambda_R} {3} \sigma_{SB} \nabla T_r^4  = -   \frac{16 \sigma_{SB} } {3 \rho \kappa_R} T_r^3 \nabla Tr \equiv - \kappa_r \nabla T_r
\end{equation}
where the Rossland transport mean free path is $\lambda_R \approx 1 / ( \rho \kappa_R )$ and $\kappa_R$ is the Rossland averaged opacity.  The Rossland opacity is based on a simple fit approximating the tabular opacity data \cite{vold2010dtburn}.

\subsection{Temperature Coupling Coefficients}

Ion and electron energy is exchanged through ion-electron collisions \cite{trubnikov, dolan1982fusion}. Electron-radiation coupling terms represent energy exchange between  the  plasma  and radiation  by electron-photon Compton scattering and by a Planck opacity absorption and emission.  The coupling coefficients used here were developed in a simple model \cite{vold2010dtburn} to study a DT run-away burn problem in an infinite space \cite{MolvigAlme}.  It was found this simple model agrees with analysis and with more sophisticated grey-body diffusion computational models \cite{vold2010dtburn}.

\textit{Ion-Electron Coupling.}
The ion-electron coupling term characterizes Coulomb collisions and  assumes ion and electron distributions are near Maxwellian \cite{trubnikov}.  The ion-electron energy exchange can be characterized by an energy equilibration time for each ion species \cite{dolan1982fusion},

\begin{equation}\label{5}
\tau_{ie} = \frac{3(2 \pi^{3})^{1/2} \epsilon_{0}^{2} m_{i} m_{e}}{n_{e} q_{i}^{2} q_{e}^{2} L} \left( \frac{kT_{e}}{m_{e}} + \frac{kT_{i}}{m_{i}}\right)^{3/2}
\end{equation}
where $\epsilon_{0}$ is the permittivity of free space, $m_{i}$ and $m_{e}$ are the masses of the ions and electrons respectively, $q_{i}$ and $q_{e}$  are the charges of the ions and electrons, and L is the Coulomb logarithm. 
The ion-electron coupling coefficient $\omega_{ie}$ is then given by 

\begin{equation}
\label{6}
\omega_{ie} = \frac{n_{i}}{\tau_{ie}}
\end{equation}
where $n_{i}$ is the ion number density, and the ion contributions are summed over ion species.

\textit{Radiation Coupling-Compton Scattering.}
The Compton scatter coupling term $\omega_{c}$ represents the transfer of energy from radiation to electrons through photon on free electron collisions. The coupling term is given in a grey-body approximation by:

\begin{equation}\label{7}
\omega_{c} = \frac{ 16 \sigma_{T} n_{e} \sigma_{SB}}{m_{e} c^{2}} T_{r}^{4}
\end{equation}
where $\sigma_{T}$ is the Thomson cross-section, $\sigma_{SB}$ is the Stefan-Boltzann constant, $m_{e}$ is the mass of the electron, $n_{e}$ is the atomic density of electrons, and c is the speed of light.

\textit{Radiation Coupling-Planck Opacity.}
The Planck opacity coupling coefficient $\omega_{p}$ determines the rate at which radiation couples to the electrons by emission and absorption processes. 
The linearized Planck coupling coefficient is  calculated as:

\begin{equation}\label{8}
\omega_{p} = 4 \rho K_{p} \sigma_{SB} (T_{e}^{2} + T_{r}^{2}) (T_{e} + T_{r}).
\end{equation}
where the Planck opacity, $K_p$,  is evaluated as a power law fit  \cite{vold2010dtburn} to tabulated values from the SESAME data base at LANL and applied here to the  deuterium fuel.

\subsection{Viscous Stress Tensor in Spherical Coordinates}

The viscous stress tensor was originally evaluated  as $(\eta_o \nabla \vec{u} ) \nabla \cdot \vec{u}$ as in Eq.~(\ref{Ti_eqn}) however this was re-evaluated in spherical coordinates to simplify in 1D with the implied symmetries, $u_{\theta} = u_{\phi} = 0$ and $\partial / \partial \theta =  \partial / \partial \phi = 0$.   The radial stress as a function of the symmetric strain components becomes

\begin{equation}
( \nabla :  \tau )_r [1 D] = - \frac{ 1 } { r ^2} \frac{\partial \tau_{rr} }{\partial r}   + \frac{\tau_{\theta \theta} + \tau_{\phi \phi} } {r}
\end{equation}
 Simplifying the stress components by the symmetry assumptions, then
$\tau_{rr} = -2 \eta_o (\partial u /  \partial r ) + (2/3) \eta_o \nabla \cdot u$ and
 $\tau_{\theta \theta} = \tau_{\phi \phi}  = -2 \eta_o  ( u /  r  ) + 2/3 \eta_o \nabla \cdot u$, with $u$,  the radial velocity component,  the tensor radial component becomes,

\begin{equation}
( \nabla :  \tau )_r [1 D] =  \frac{ 1 } { r ^2} \frac{\partial}{\partial r}
 \left( r^2 2 \eta_o \left(  \frac{\partial u}{\partial r} + \frac{\nabla \cdot u}{3} \right) \right)   
 - \frac{4 \eta_o } {r}\left( \frac{u}{r} - \frac{\nabla \cdot u}{3} \right)
\label{visceqn1}
\end{equation}
These four terms are interpreted as a diffusion of velocity, a compressible correction to the diffusion, a spherical metrics correction and the compressible contribution to the spherical coordinate correction. 
Evaluating the divergence in spherical coordinates, $\nabla \cdot u = (1/r^2) \partial ( r^2 u ) /  \partial r  = \partial u / \partial r + 2 u / r$, the terms can be rearranged to simplify as

\begin{equation}
( \nabla :  \tau )_r [1 D] =  \frac{ 1 } { r ^2} \frac{\partial}{\partial r}
 \left( r^2 \frac{4}{3} \eta_o \left(  \frac{\partial u}{\partial r} - \frac{u}{r} \right)  \right)  
 + \frac{4 \eta_o } {3 r}\left( \frac{\partial u}{\partial r} - \frac{u}{r}  \right)  
\end{equation}

The form for viscous dissipation in the energy equation with the assumed radial symmetry, $[ u_{\theta} = u_{\phi} =  \partial / \partial \theta =  \partial / \partial \phi = 0 ]$, becomes
\begin{equation}
\Phi_{visc} = 
 \eta_o \left( 2 \left( \left( \frac{\partial u}{\partial r} \right)^2 + 2  \left( \frac{ u}{ r} \right)^2  \right) 
- \frac{2}{3} \left( \frac{1}{r^2} \frac{\partial}{\partial r} ( r^2 u )\right)^2  \right)
\end{equation}

Numerical issues associated with the discretization of the viscous terms are discussed briefly in the next section.

\section{Numerical Methodology}

The numerical scheme is based on a Lagrange velocity evaluated at mesh cell faces or nodes, i,  located between thermodynamic quantity cell centers indexed as $i$ and $i-1$.  At initial times a grid  with $\Delta r \approx 2 - 3 \mu m$ allows the Lagrange cell sizes to remain manageable ($\Delta r_{min} \approx 0.3 \mu m$) at the maximum compressions in this study.  With the pressures and  artificial viscosity \cite{wilkins}, $q_i$ , evaluated at cells, $i$, and $i-1$, a new (advanced time) Lagrange velocity, $u_i^{n+1}$ is first determined at velocity node, $i$.  This is used to calculate the new  position of each Lagrangian co-ordinate and advance the density, momentum and temperatures in the Lagrange frame.
\begin{align}
q_i &= 
\begin{cases}
\rho_i |\Delta u_i| (c_Q |\Delta u_i|+c_L c_{s,i}), & \text{if } \Delta u_i < 0 \\
0, & \text{otherwise}, \end{cases} \\
u_i^{{n+1}} &= u_i^n - \frac{\Delta t}{\rho \Delta r}((P_{\text{all}}+q)_i^n - (P_{\text{all}}+q)_{i-1}^n) \label{ueq} \\
u_{\text{all}}^{n+1} &= \nabla \cdot \eta_0 \frac{\Delta u_{\text{all}}^n}{\Delta r} \label{viscdiff}\\
r_i^{n+1} &= r_i^n - \frac{1}{2}\Delta t (u_i^{n+1} + u_{i}^n) \label{req} \\
\rho_i^{n+1} &= \rho_i^n \frac{V_i^n}{V_i^{n+1}} \\
\end{align}

The three temperature equations are then solved explicitly  for the viscous terms and for half the PdV work done by pressure on each species, 

\begin{align}
Ti_i^{n+1} &= Ti_i^{n} - \frac{dt}{dr_i} ~  (\gamma - 1) ~ \frac{\text{Vol}_i^{n+1}}{N_{i~\text{ions}}} ~  \frac{0.5 Pi_i + q_i - \Phi_{visc} }{r_{i+1}^2 u_{i+1}^{n+1} - r_{i}^2 u_{i}^{n+1}} \\ 
Te_i^{n+1} &= Te_i^{n} - \frac{dt}{dr_i}  ~ (\gamma - 1) ~ \frac{\text{Vol}_i^{n+1}}{N_{i~\text{electrons}}} ~ \frac{0.5 Pe_i}{r_{i+1}^2 u_{i+1}^{n+1} - r_{i}^2 u_{i}^{n+1}} \\ 
Tr_i^{n+1} &= Tr_i^{n} - \frac{dt}{dr_i} ~  \frac{c}{16 \sigma_{SB} Tr_i^3} ~ \frac{0.5 Pr_i} {r_{i+1}^2 u_{i+1}^{n+1} - r_{i}^2 u_{i}^{n+1}} 
\end{align}
The three temperatures are then updated  implicitly for the remaining half of the PdV work, and lastly for the heat conduction of each species denoted as,
\begin{align}
Ti_{\text{all}}^{n+1} &=  \nabla \cdot \kappa_i \frac{\Delta Ti_{\text{all}}^n}{\Delta r}  \label{Tidiff}\\
Te_{\text{all}}^{n+1} &=  \nabla \cdot \kappa_e \frac{\Delta Te_{\text{all}}^n}{\Delta r}  \label{Tediff} \\
Tr_{\text{all}}^{n+1} &=  \nabla \cdot {\kappa_r \frac{\Delta Tr_{\text{all}}^n}{\Delta r}}   \label{Trdiff}
\end{align}

The implicit differencing of the diffusion operators in the computations, Eq. (\ref{viscdiff}, \ref{Tidiff}, \ref{Tediff}, \ref{Trdiff}) are performed using the Thomas algorithm  solving a tridiagonal matrix. 
Following Eq. (\ref{Trdiff}), the three temperature coupling is performed through an iterative procedure until convergence for each of the three temperatures is achieved. 

After the three temperature equations are evaluated, the species mass mixing algorithm is executed. The code has the option of performing mass species mixing implicitly when using only the molar gradient term in Eq. (\ref{massmix}) or calculating mix explicitly if  other terms  are desired. The species mass transport coefficient, $\mu = (p_{ia} / \rho ) (\rho_j / \nu_{ij})$ is  computed as the harmonic mean of the cell centered values adjacent to the interface of interest.  The $dr$ factor from the gradients in each term is also included in the harmonic mean to account for the time varying Lagrange geometry of the cells.

\begin{align}
 \left(\frac{p_{ia} \rho_j}{\nu_{ij}\rho}\frac{1}{dr} \right)_{\pm 1/2} = 2.0\left(\left(\frac{p_{ia} \rho_j}{\nu_{ij}\rho}\frac{1}{dr} \right)^{-1}_{i\pm1}+\left(\frac{p_{ia} \rho_j}{\nu_{ij}\rho} \frac{1}{dr} \right)^{-1}_{i}\right)^{-1}
\end{align}

It is noted that $\rho_j/\nu_{ij}$ results in the cancellation of the density of the heavier species,  $n_j$, and thus avoids dividing by zero, where the heavy ion molar density, $n_j$  is zero.

The final step  of the hydrodynamics at a new time is the computation of the  pressures for the ions, electrons, and radiation, $P_i, P_e, P_r$.  These are given as functions of the new values for mix species concentrations and temperature at each cell,

\begin{align}
P_i^{n+1} &= (n_{DD}^{n+1} + n_{CH}^{n+1}) k_B T_i^{n+1} \label{ionpress} \\
P_e^{n+1} &= n_e^{n+1} k_B T_e^{n+1} \label{elecpress} \\
P_r^{n+1} &= \frac{4}{3} \frac{\sigma_{SB}}{c} (T_r^{n+1})^4 \label{radpress} 
\end{align}
where $n_{DD}$ and $n_{CH}$ refer to the ion number densities for the deuterium and for the CH.  The electron density is a function of the ionization and ion density at the specified grid point, $n_e = (z_{DD} n_{DD}) + (z_{CH} n_{CH})$ where $z$ is the ionization, and is set to the atomic charge  when the first shock passes.\\\

\textit{Viscous Terms.}
A centered difference approximation for the first term in Eq. (\ref{visceqn1}) leads to the standard diffusion discretization, which is implicitly stable or can be explicitly stabilized with a  time step restriction of order, $\Delta t \le \Delta r^2 / D$, where $D \approx \eta / \rho$.  Numerical stability  occurs when the   discretization of the viscosity in the momentum equation leads to a negative coefficient for  the radial velocity  at the centered node, $i$,  which in turn  leads to a positive contribution to the diagonal coefficient in the implicit matrix solver for the viscosity.  This stabilization occurs naturally for the  fourth term in the viscous tensor, $ - (4\mu / 3 r) ( u /  r)$.  The second and third  terms could be stabilized by a type of 'upwind' differencing, where the difference approximation for the $\partial u = \Delta u$ is written as $ \Delta u \approx u[i] - u[i-1] $ or $\Delta u \approx u[i+1] - u[i] $, chosen to retain a negative coefficient for $u[i]$.  However, this introduces numerical dissipation, and we have found the second and third terms can be center-differenced for higher order accuracy and are well behaved in an implicit solution method.  When solved implicitly these new terms modify the off-diagonal coefficients without modifying the diagonal.

For the cell centered viscous energy dissipation in the standard Lagrangian staggered grid, the velocities and radii values at cell faces allow a straight forward difference approximation to the velocity gradients evaluated in the cells.  The term, $(u/r)$, is evaluated as an average of the cell's face values.

\section{Results}

\textit{Basic Tests.}
The hydrodynamic and plasma physics models underwent extensive testing first as a one temperature code \cite{FennMollVold} and then in the 3T version.   The code recovers the expected solutions for several standard problems (e.g., Sod shock tube, Guderley spherical test, planar diffusion).   Grid convergence tests indicate as expected that the inviscid solutions do not converge in the zones at the center of symmetry for shock solutions imploding in spherical geometry.  However, the dissipation introduced by the plasma transport terms allows a converged solution for the central peak temperature for grids with at least 200 zones in the ICF problem \cite{FennMollVold}.  The plasma thermal conduction is the most important term for grid convergence of the temperature but plasma viscosity also plays a signficant role.    The 3T temperature model was extensively tested \cite{vold2010dtburn}  against other codes and in analysis for a test problem representing a DT runaway burn problem in an infinite space \cite{MolvigAlme}.

\textit{The base case.} An ICF implosion of a plastic, CH, capsule of 25 micron thickness with inner radius of 400 microns imploding on a DD fuel was selected and used to validate the 1D code model and algorithms. The laser  power is 22 TW with an absorption fraction (default set to 60\%)  delivered to the electrons in the outer CH zones over a 1 ns square pulse assumed to be representative of an Omega-scale ICF implosion.  Setup details were previously described \cite{FennMollVold} and ICF test results from the code were found to be in reasonable  agreement  with the HELIOS ICF code \cite{Macfarlane, welser1, welser2}.  The binary mix model assumes average ions in each material, with $z_{CH} = 3.5$ and $A_{CH} = 6.5$ in the CH plastic capsule.

\subsection{Comparison between Inviscid and Viscous Plasma Transport}

\begin{figure}[H]
\subfloat[]
{\includegraphics[width=0.9\textwidth]{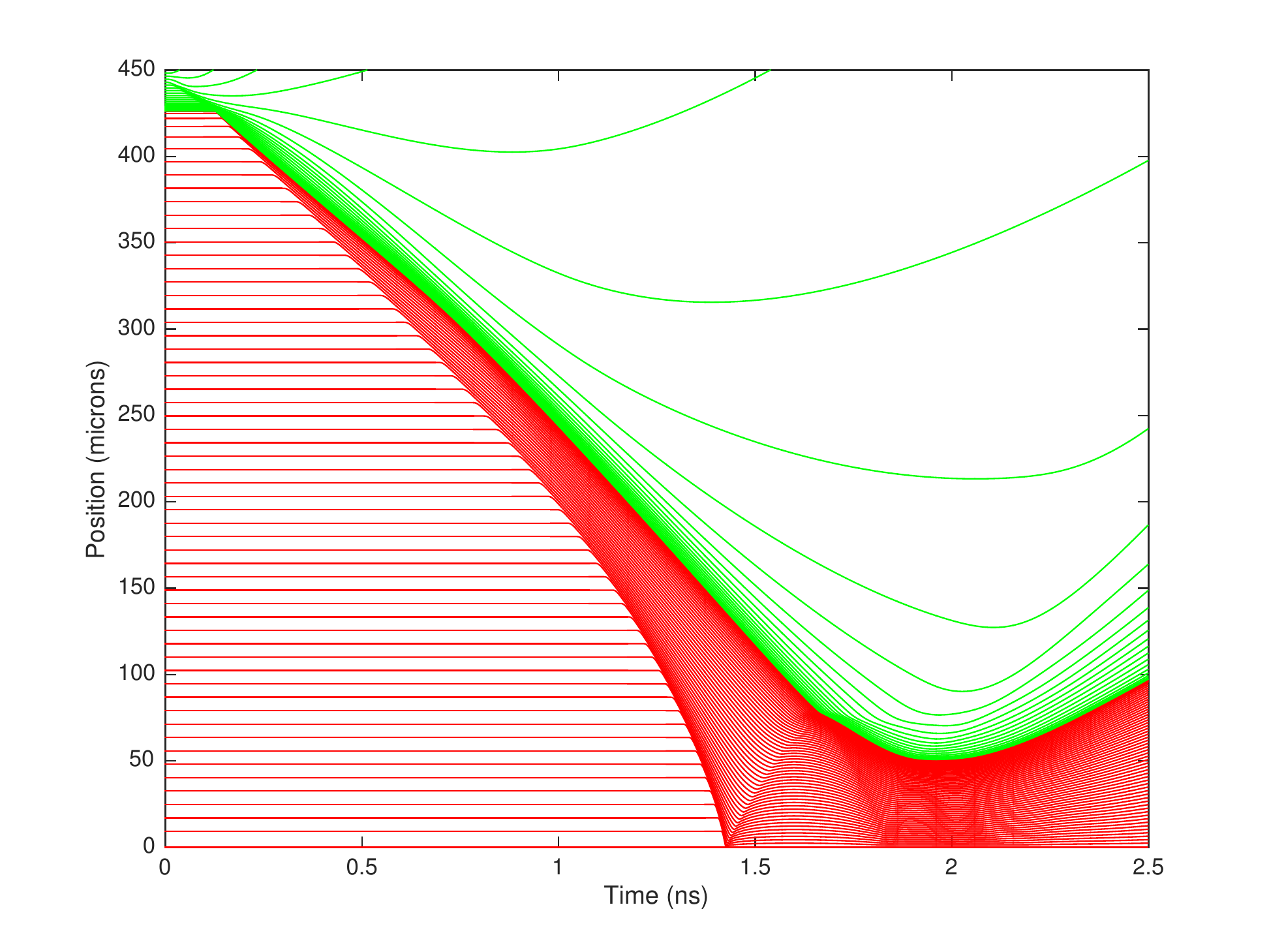}}
\label{rtbase}
\subfloat[]
{\includegraphics[width=0.9\textwidth]{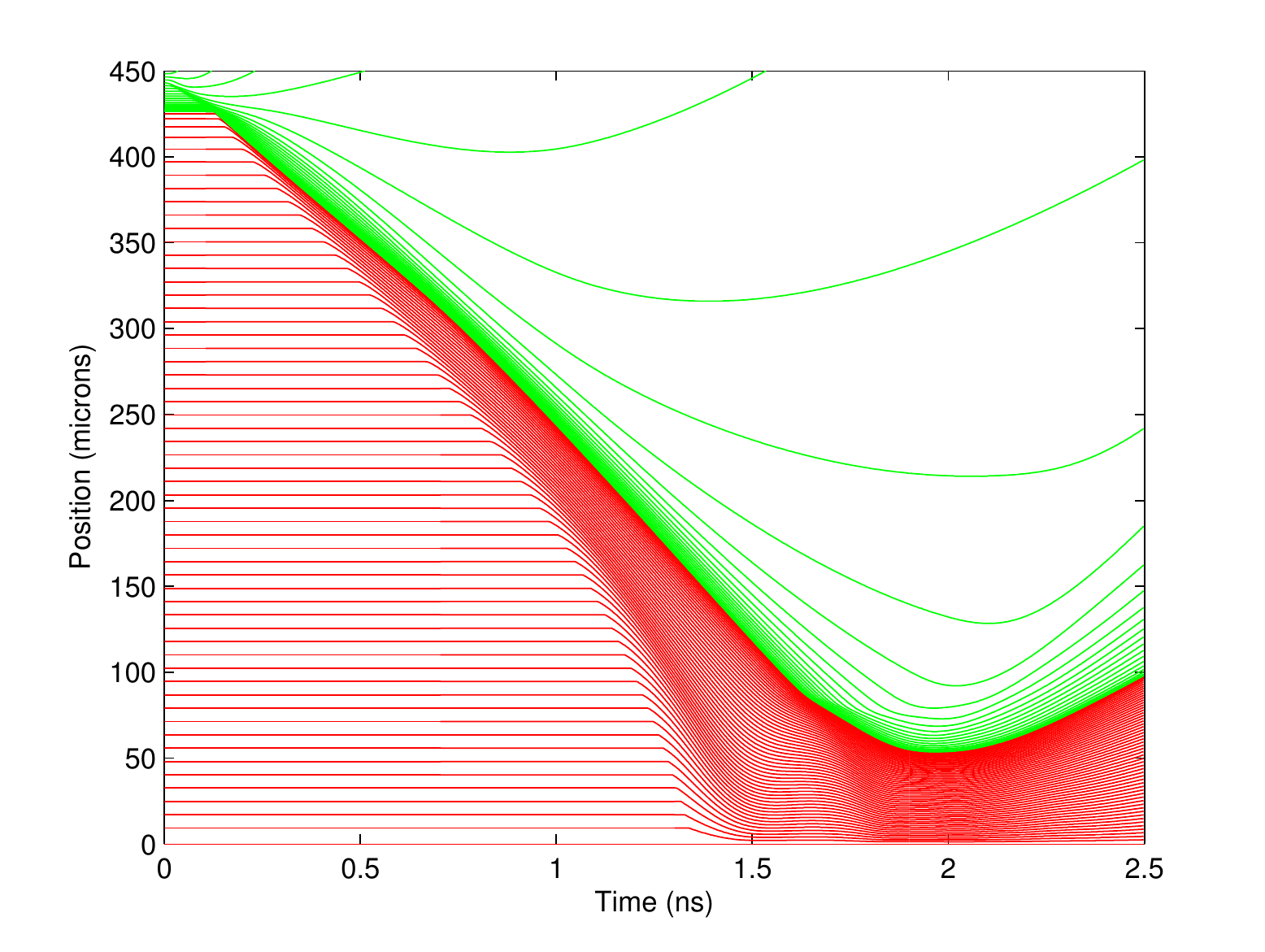}}
\caption{(a) Radius-Time Plot for ICF Base Case without viscosity.  (b) Radius-Time plot of the same case with viscosity. The Red and Green colors refer to fuel and plastic capsule, respectively.}	
\label{rtcmp}	
\end{figure}

A radius-time plot of the Lagrangian fuel and plastic zones for the base case is shown in Fig. \ref{rtcmp} for the inviscid (a) and for the viscous (b) solutions. Each line represents a Lagrangian coordinate  showing the movement and compression of the fuel and capsule. Laser heating of the plastic zones can be seen at the beginning of the simulation and the compression of the fuel continues until it is approximately 55 - 60 microns in radius. The shock first arrives on center between 1.4 to 1.5 ns and the return shock arrives at the inner capsule surface at approximately 1.7 ns, while the time of maximum compression is $\approx$ 1.9 ns.

The incoming shock front is sharp in the inviscid case and remains steep after convergence on center.  The first and second reflections at the fuel-capsule interface are apparent in the figure.  The viscous shock has a less distinct front location as it converges on center and after reflection it becomes even more indistinct.  The difference in the timing of incidence of the main shock is approximately $0.1$ ns. The inviscid shock arrives at  1.4 ns and returns to the center point at  1.8 ns. The viscous shock arrives at about 1.5 ns with a leading  'toe' blurring the main shock front by about 0.1 ns.  This shock also returns to the center point at about 1.8 ns, when it appears to be much more diffuse than the inviscid case.

As seen in  Fig.~\ref{rtcmp}, the fuel reaches a minimum radius of approximately 50 - 60 $\mu$m for the inviscid and viscious cases  
The maximum compression is less for the viscous case, although the difference appears to be  small in the radius-time plots.  This will be examined in more detail when comparing the position of the composition fraction profiles for the fuel-capsule mixing layer.

As the first shock is reflected from the center and returns to the capsule inner shell the inviscid shock is clearly sharper in the fuel and this creates a steeper shock rebounding into the plastic compared to the viscous case.  However, Figure \ref{rtcmp} suggests that viscous effects do not greatly affect the plastic.   As the   laser deposited energy propagates  into the fuel, the plastic temperature decreases, and  the strong temperature dependence of the ion viscosity reduces its role in the plastic.  The higher average atomic mass of the plastic and especially the greater average ionization state also reduces the viscosity in the plastic.  \\

\begin{figure}[H]
\begin{center}
 \includegraphics[width=0.75\textwidth]{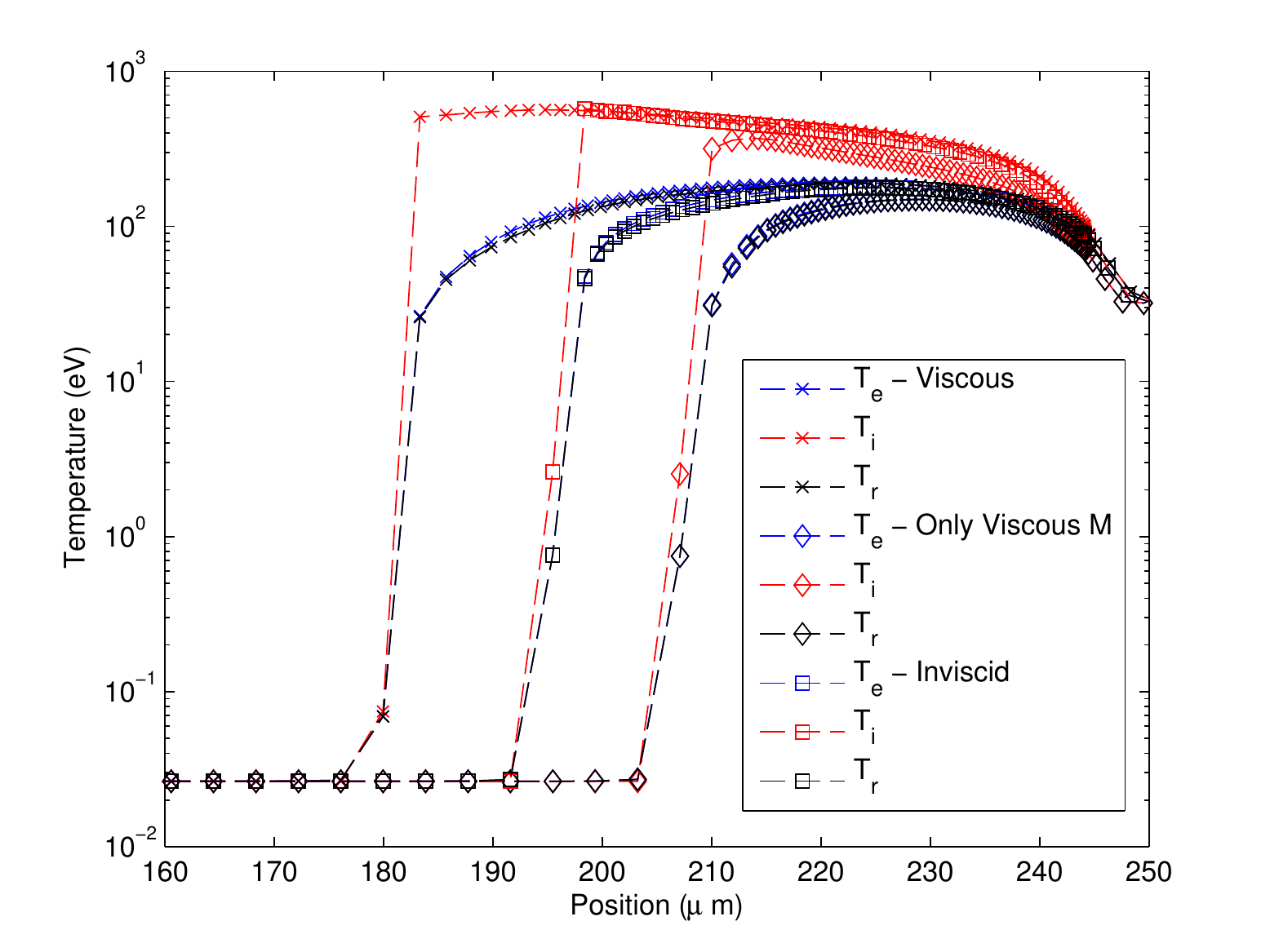}
\caption{Electron, Ion, and Radiation Temperature profiles for 3 cases: Inviscid, Viscous, and viscosity in momentum equation only  labelled 'Only Viscous M'  (no viscous energy dissipation) at $t = 1.0$ ns.}
\label{cmp2allT}
\end{center}
\end{figure}

A snapshot of  three temperature profiles at 1.0 ns, are shown in Fig.~\ref{cmp2allT} for three cases: viscous, inviscid and momentum viscosity only (with no viscous dissipation in the energy equation). At this early time the shock fronts are near a radius of $200 \mu m$, about half the initial fuel radius, so that convergence effects are still minimal.  The electrons and radiation temperatures are nearly identical in each case while the ions are shocked to a higher temperature with a steep front in each case, and there is evidence of a small 'toe' forming at the shock front in the viscous case.  Electron thermal conductivity diffuses the electron temperature behind the shock front, and the temperatures are still small enough at this time that the electron temperature has  not propagated  beyond the ion shock front.  Since the thermal conductivity for the ions is smaller by a factor of $\sqrt{m_e/m_i}$, the ion shock front lacks the diffusive characteristic.  

The viscous shock leads the inviscid shock at this time by about 15 cm or about 7\% of the radius while the shock in the third case, with momentum viscosity only and no viscous dissipation in the energy equation, lags the inviscid case by about the same amount.  The differences in shock positions are attributed primarily  to shock energy differences with the heat from the viscous dissipation increasing the energy in the shock.  With no viscous heating in the temperature equation, the viscosity in the momentum equation dissipates the shock velocity slowing the shock front compared to the inviscid case.


A point of interest arises in comparing  the shock fronts at 1 ns, in Fig.~2  to the shock fronts and arrival time at the target center as seen in the radius-time plots in Fig.~1.  Clearly, the viscous shock leads the inviscid shock by $\approx$ 15 microns or 7\% of the radius at 1 ns.  However, once the shocks converge on center as seen in Fig. 1, the shock with viscosity has blurred and the main front apparently  reaches the center at about 1.5 ns,  lagging the inviscid case arriving on center at 1.4 ns.   This suggests that the viscous effects vary significantly with the radial convergence.  When there is little convergence at 1 ns the viscous shock leads the inviscid while the shock dissipation by viscous effects is enhanced in the convergent geometry, especially near the center point, leading to the viscous shock front lagging the inviscid shock during convergence at the center.\\\

\begin{figure}[H]
\subfloat[]
{\includegraphics[width=0.5\textwidth]{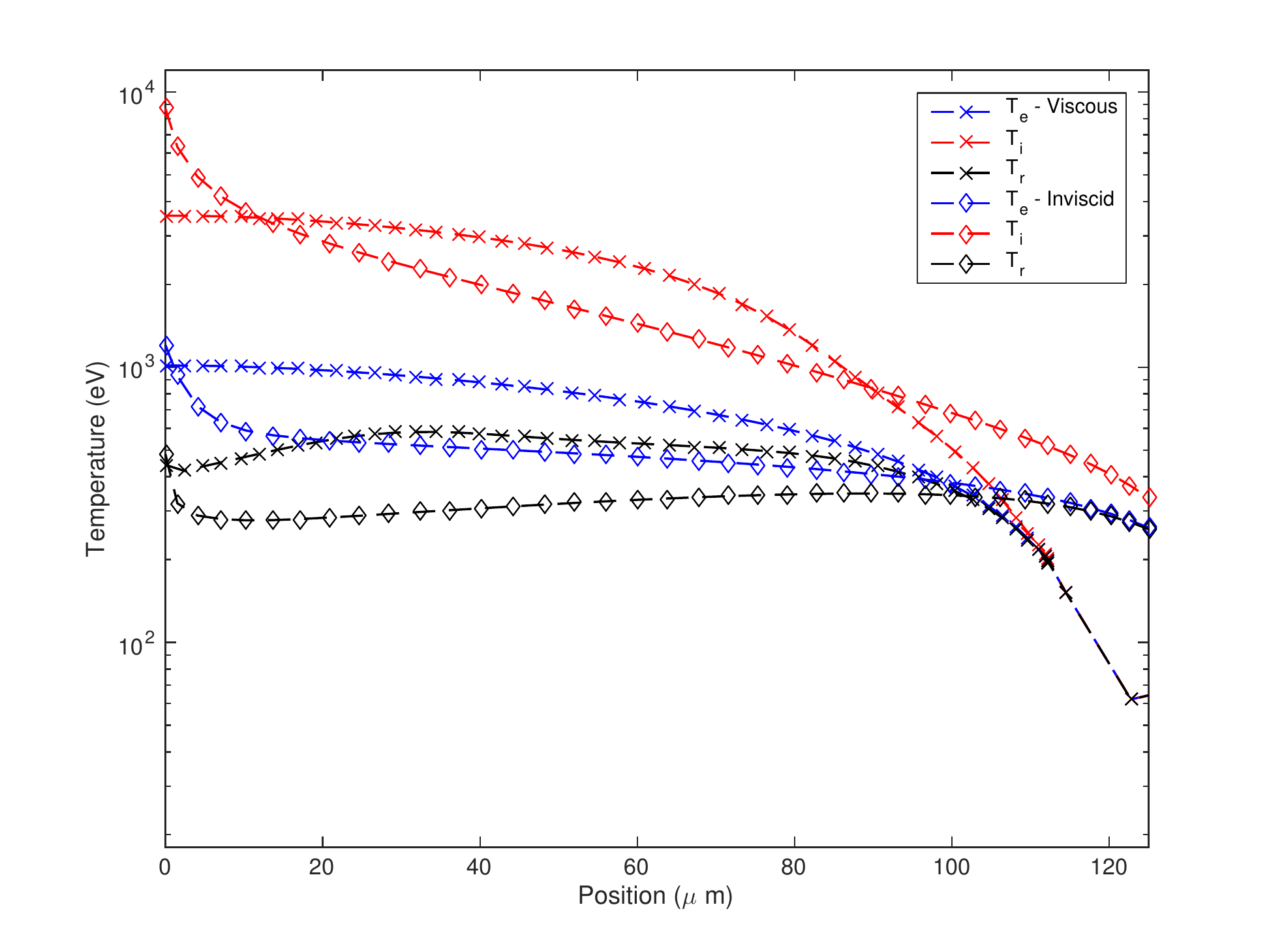}}
\subfloat[]
{\includegraphics[width=0.5\textwidth]{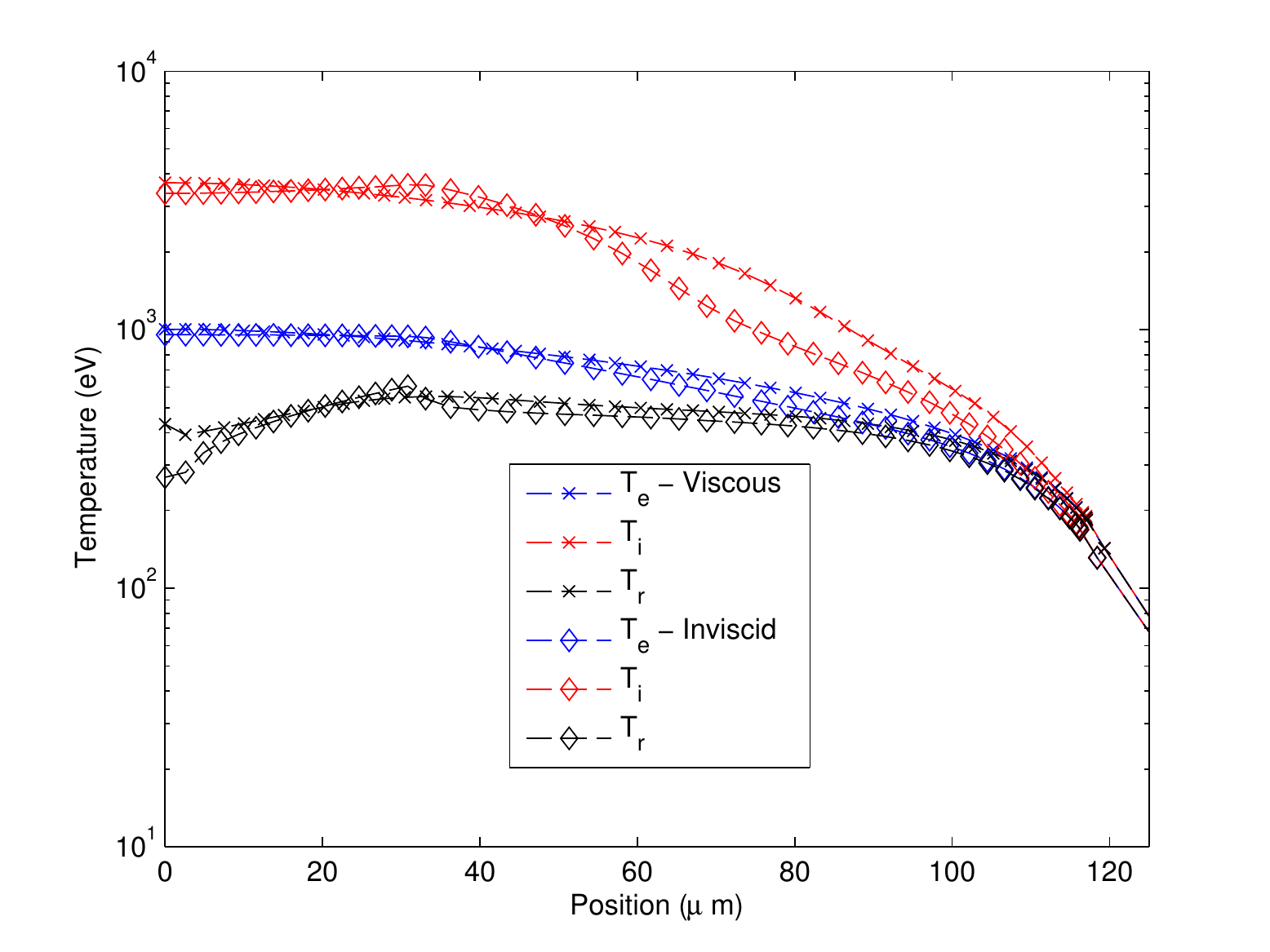}}
\caption{Ion, Electron, and Radiation Temperatures at times (a) of first shock convergence and (b) just after first shock convergence.}
\label{3Tbase}		
\end{figure}

A portion of the radial profiles of the  temperatures are shown in Fig.~\ref{3Tbase}(a) during shock convergence on the target center for  the viscous and the invisicid cases, respectively at  times of 1.52 ns and 1.42 ns. The inviscid and viscous profiles are  compared  at a  time  shortly after convergence (1.56 ns  for each case) in Fig.~\ref{3Tbase}(b).
Considering first the time of shock convergence in Fig.~\ref{3Tbase}(a),  the inviscid case shows a strong increase  in temperatures at the radial center driven by the  shock heating of the ions.  At this time the viscous case shows no significant local increase in temperatures  apparently due to the viscous dissipation.  The ion-electron and electron-radiation temperature differences are significant and the differences are comparable between the viscid or inviscid cases.  Immediately after the shock convergence in Fig.~\ref{3Tbase}(b), the  central temperatures relax in the inviscid case to the values seen in the viscous case.

\begin{figure}[H]
\begin{center}
{\includegraphics[width=0.5\textwidth]{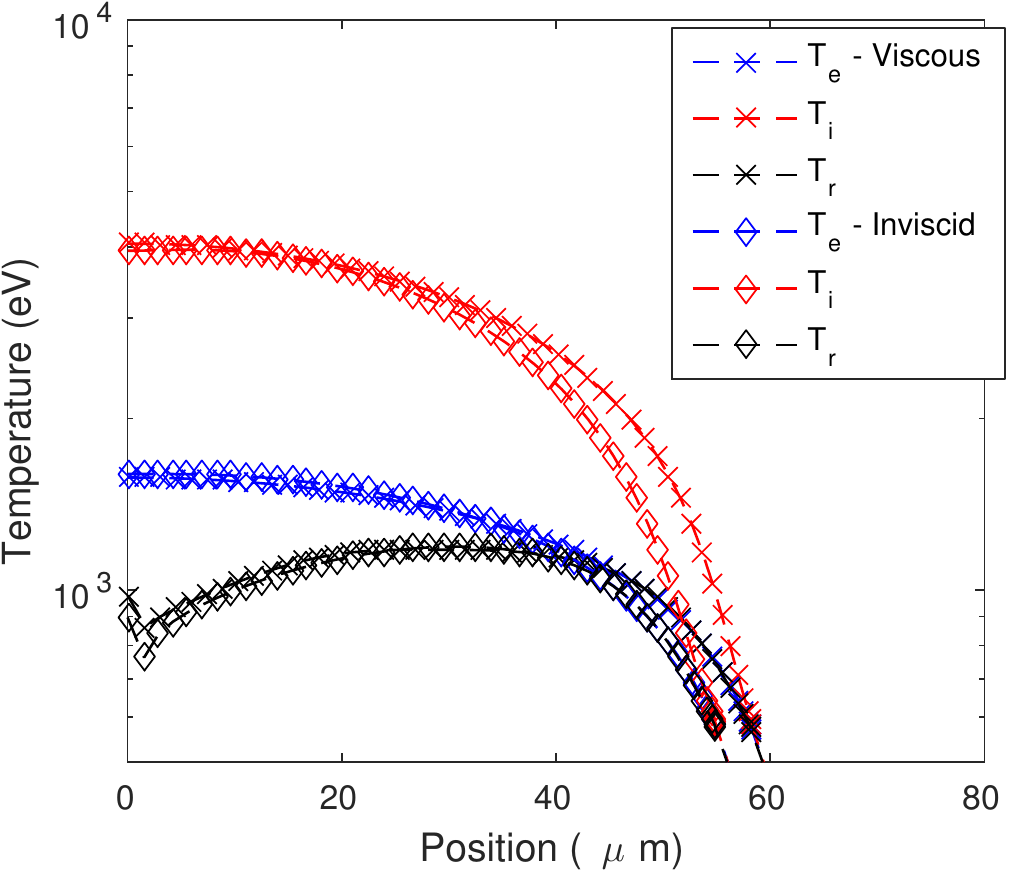}}
\caption{Ion, Electron, and Radiation Temperatures at times  near maximum compression and neutron production.}
\label{3Tbase2}
\end{center}
\end{figure}


Temperature profiles in Fig.~\ref{3Tbase2}, are shown at a later time near maximum compression, ($t = 1.85 ns$ for  both the inviscid and viscous  cases, the time of maximum neutron production).  These profiles are similar in shape to that seen immediately after the shock converged, although compressed in radius and slightly elevated in maximum temperature values on center.   The viscous and inviscid profiles are very similar suggesting the burn dynamics will be similar as discussed next.  At each of the times, in Fig.~\ref{3Tbase} and  Fig.~\ref{3Tbase2}  there are  large temperature differences between  $T_i, T_e, T_r$, suggesting the 3T model is  important  for the inviscid or viscous case.    The radiation energy density is small but the ion and electron temperatures are influenced by  the radiation through the temperature coupling terms described previously.   The significant temperature differences at these times are in contrast to the temperatures at  the early implosion time, $t = 1 ns$ seen in Fig.~\ref{cmp2allT}, when the electron and radiation temperatures are tightly coupled, and only the ion temperature is significantly greater due to the incoming shock heating.\\\

\begin{figure}[H]
\begin{center}
 \includegraphics[width=0.7\textwidth]{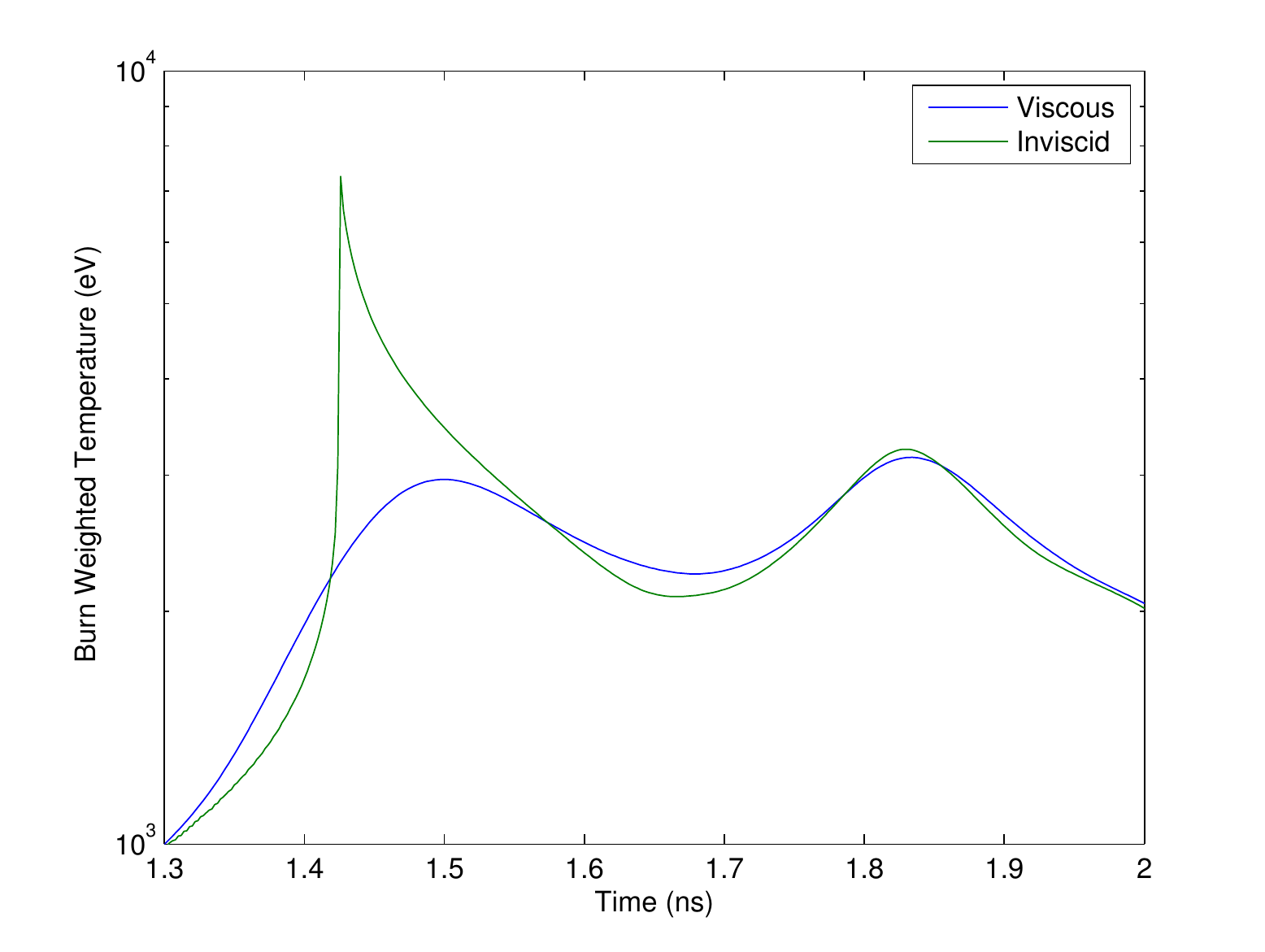}
\caption{Burn weighted ion temperature over time for both cases, viscid and inviscid. The temperature is consistently higher in the inviscid case.}
\label{cmpTw}
\end{center}
\end{figure}

The calculation of burn weighted ion temperatures, illustrated in Fig. \ref{cmpTw}, was used as a diagnostic to compare the code to experimental and other numerical results.  For each case, the time dependent  burn weighted ion temperatures exhibit two local maximum values, corresponding to the times of shock heating and compression burn. The maximum  burn weighted ion temperature in the inviscid case was found to be approximately 7 keV (and briefly approaches 10 keV at first shock heating), while the viscous case maximum burn weighted ion temperature is about 3 keV, at first shock or at maximum compression.   The difference in time dependent burn weighted temperature between viscid and inviscid cases is large only during the interval, $1.44 - 1.58 ns$, as the first shock converges and reflects from the center.  The time averaged burn weighted temperature in either case was similar, about 2.6 keV.  These values are comparable to experimental results and other 1D ICF simulations \cite{HELIOS, welser2, similarBurnandT}. \\\

\begin{figure}[H]
\begin{center}
 \includegraphics[width=0.7\textwidth]{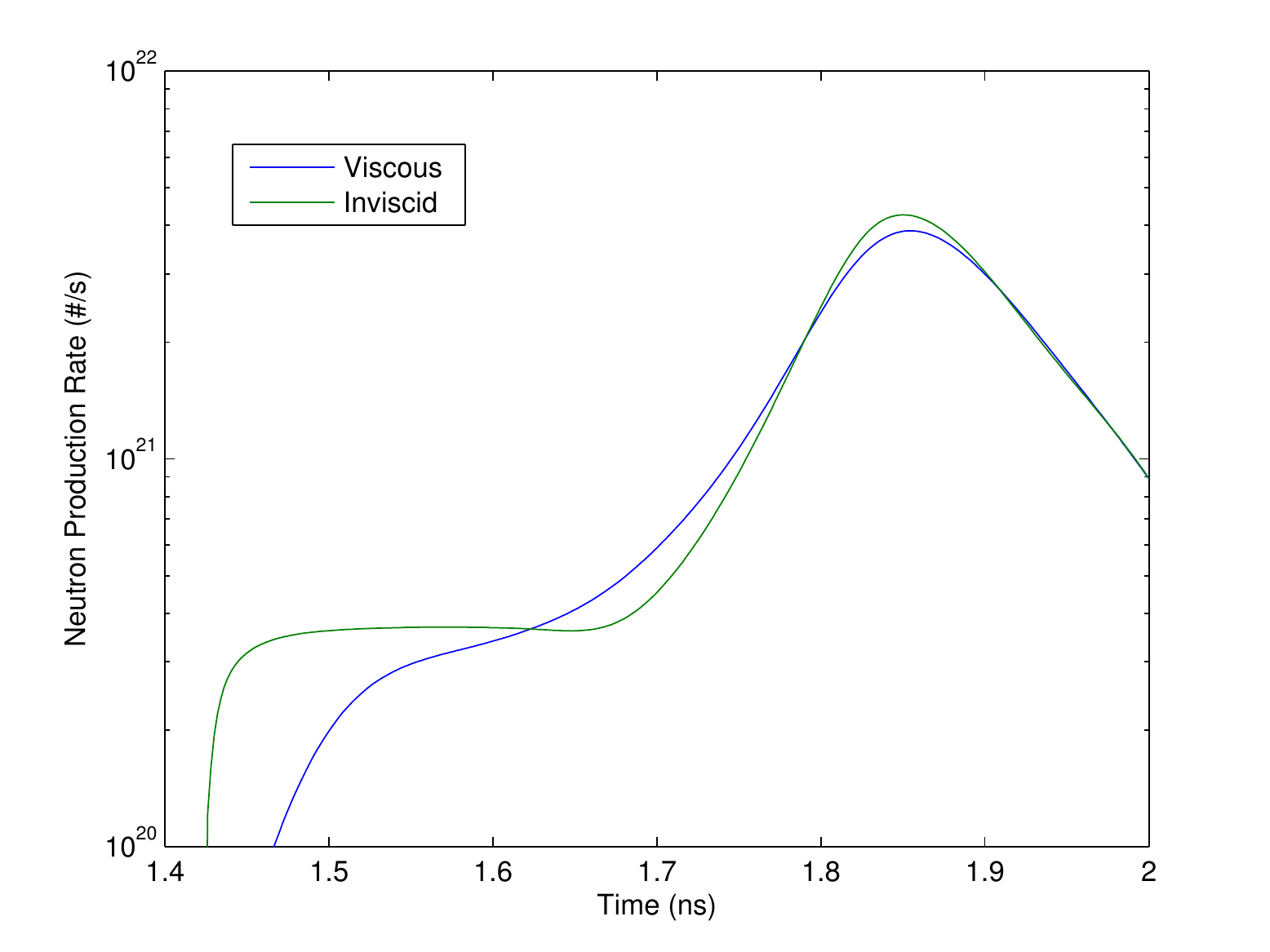}
\caption{Neutron production rate over time for both simulations, viscid and inviscid. The peak occurs at the time of maximum compression as indicated in the previous plots.}
\label{cmpburn}
\end{center}
\end{figure}

Neutron production rates verses time  in Fig. \ref{cmpburn} show yield rates and yield of order $10^{21} s^{-1} \times 0.3 ns \approx 3 \times 10^{11}$ and are comparable to experiment and previous simulations \cite{meyerhofer, Li, bradley}.  For each case, neutron production rates exhibit a local maximum value at the time of  compression burn. During shock heating,  neutron production is about 10\% of the maximum level for the inviscid case and significantly less for the viscous case. Maximum neutron production rates were found to occur approximately at 1.85 ns, which agrees well with previous studies (\cite{HELIOS, vold-sherrill, welser1, welser2}, see also\cite{bradley} and references therein).

The plots of burn weighted temperatures in Fig.~\ref{cmpTw} and the neutron production rate in Fig.~\ref{cmpburn}  illustrate the inviscid and viscous differences in shock timings. The inviscid case burn temperature has a sharp peak at the time of incidence of the first shock while the viscous  case does not since it dissipates the shock energy more effectively. The decrease in temperature indicated in Fig. \ref{cmpTw} for the viscous case leads to a delayed profile for the neutron production rate compared to the inviscid simulation.  The relatively small neutron production rate at first shock, even for the inviscid case is attributed to the lower density at that time before maximum compression.

\begin{figure}[H]
\subfloat[Inviscid]{\includegraphics[width=.45\textwidth]{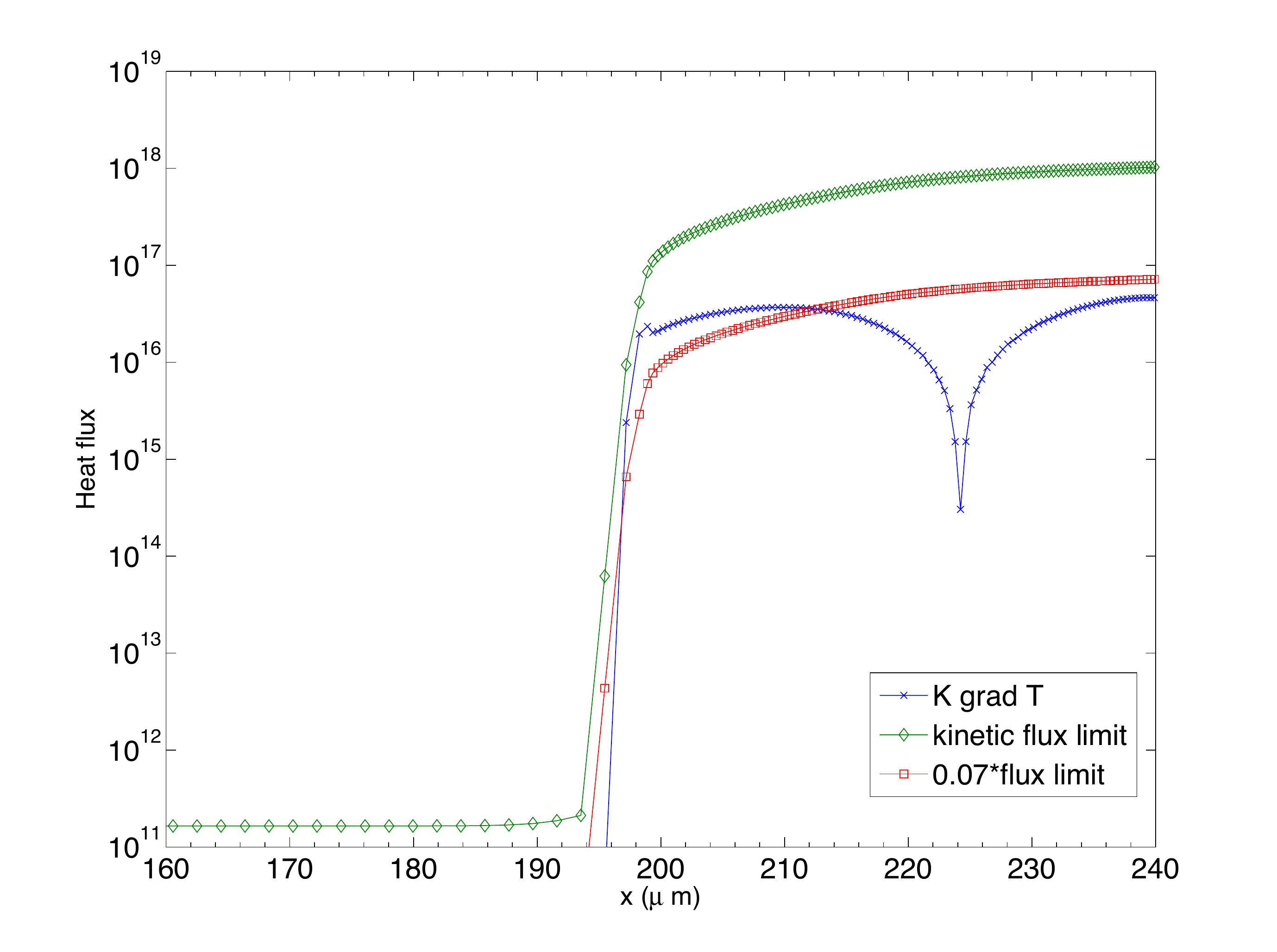}}
\qquad
\subfloat[Viscous]{\includegraphics[width=.45\textwidth]{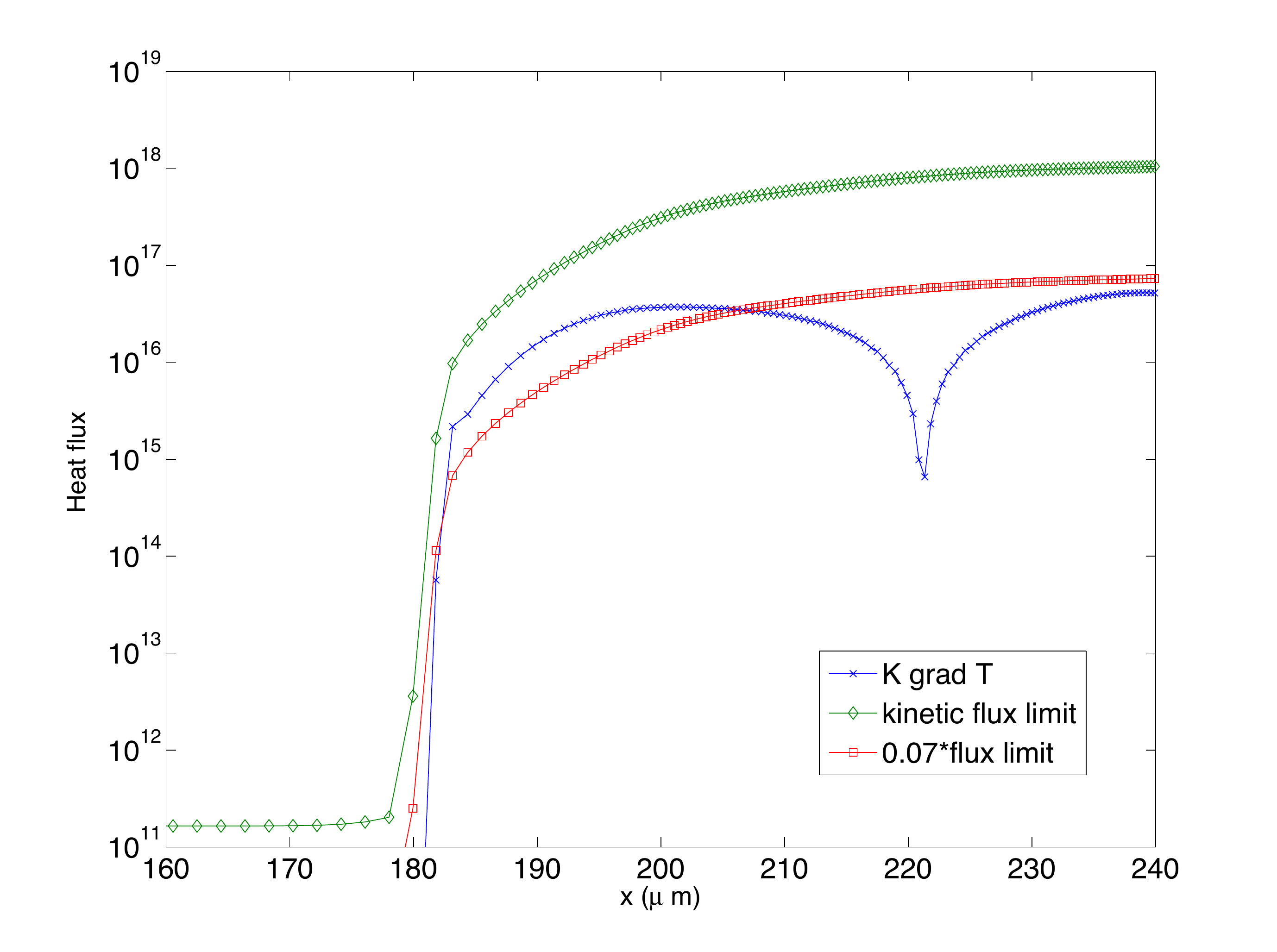}}
\caption{Plot of the heat flux, $\kappa_e \nabla T_e$, denoted (K grad T), with the kinetic flux limit, $3 n k_b T_e  v$, and the commonly implemented flux limit, $7\%$ of $3 n k_b T_e  v$}
\label{fluxl}
\end{figure}

To better match experimental results during the early stage of ICF implosions it has been proposed in previous work to use non-local thermal transport models or to use flux-limited electron transport \cite{Atzenitext, goncharov, schurtz, schurtz2007revisiting, marocchino2014effects}.  Figure \ref{fluxl} shows a plot of the classical transport electron heat flux, $\kappa_e \nabla T_e$ in our simulations, along with the heat flux at the maximum physical flux limit and with the commonly implemented flux limiter of $7\%$ of the maximum flux limit \cite{schurtz2007revisiting, marocchino2014effects}.  The maximum limited  heat flux is the magnitude of the internal energy that can be moved by a characteristic velocity, $v = \frac{2}{\sqrt{\pi}} v_{th}$.   Each is plotted at a time, $t = 1 ns$, when the laser drive has just turned off and an equal amount of laser energy has been deposited in both the viscid and inviscid simulations. 

The comparison for the viscous and inviscid cases illustrates a few points. The shape of the heat flux front is considerably steeper in the inviscid case, and the difference in the size of the shocks is  evident.  In the inviscid case, the classic heat flux exceeds the $7\%$ limited flux over the radius, $195 - 210 ~\mu m$ while  in the viscous case, this occurs over the radius $180 - 205 ~\mu m$.  These are the regions which might require the implementation of a flux limiter, and the region is $\approx 60\%$ larger in the viscous case. This suggests that viscous effects  can modify the role of a flux limiter. Where these flux limiters are implemented with phenomenological model parameters, the parameter values may need adjustment when including plasma viscous effects.  Conversely, including viscosity in simulations may reduce the need for phenomenological adjustment of the electron heat flux.  \\\

\begin{figure}[H]
\subfloat[Inviscid]{\includegraphics[width=.45 \textwidth]{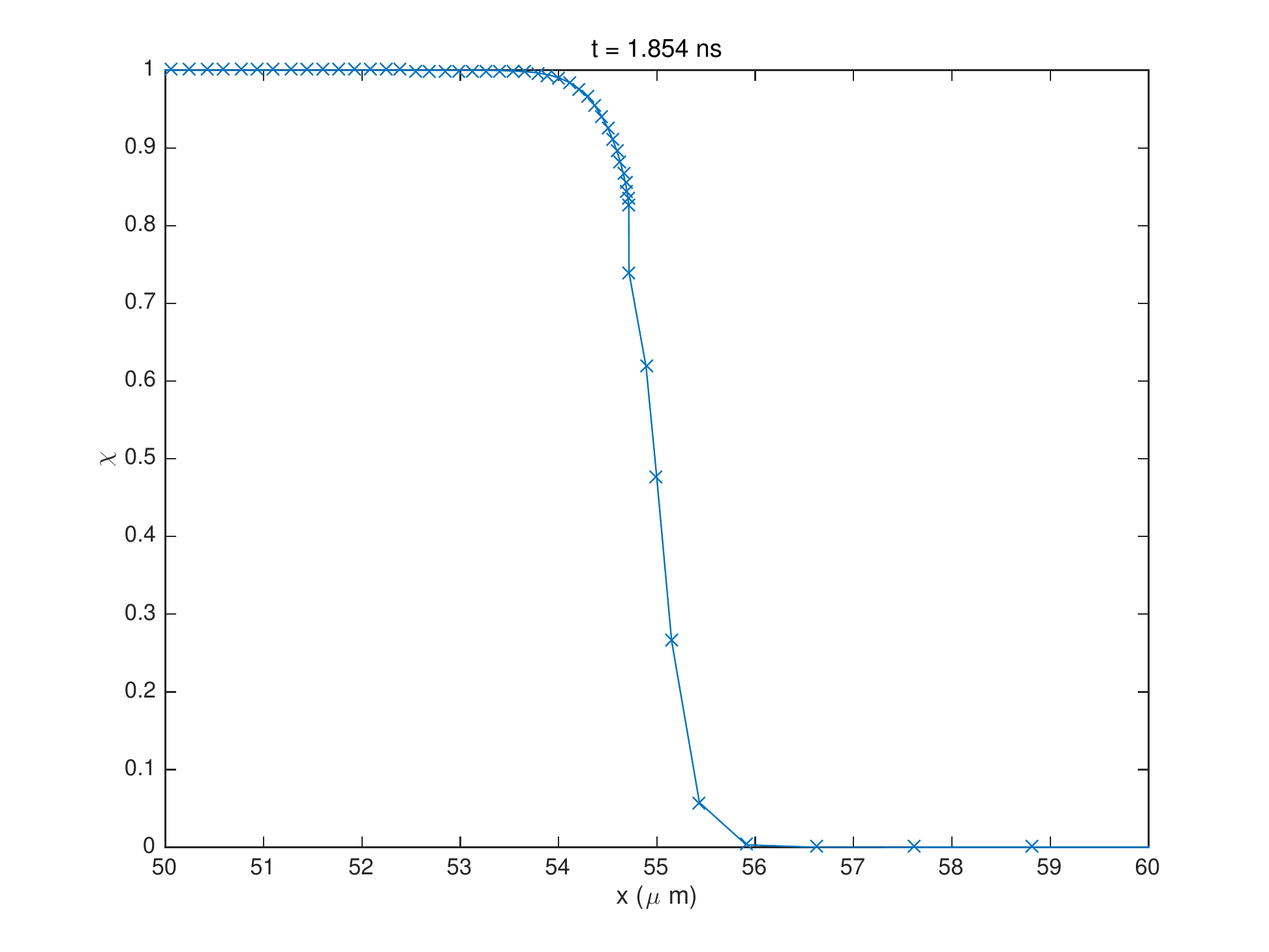}}
\qquad
\subfloat[Viscous]{\includegraphics[width=.45\textwidth]{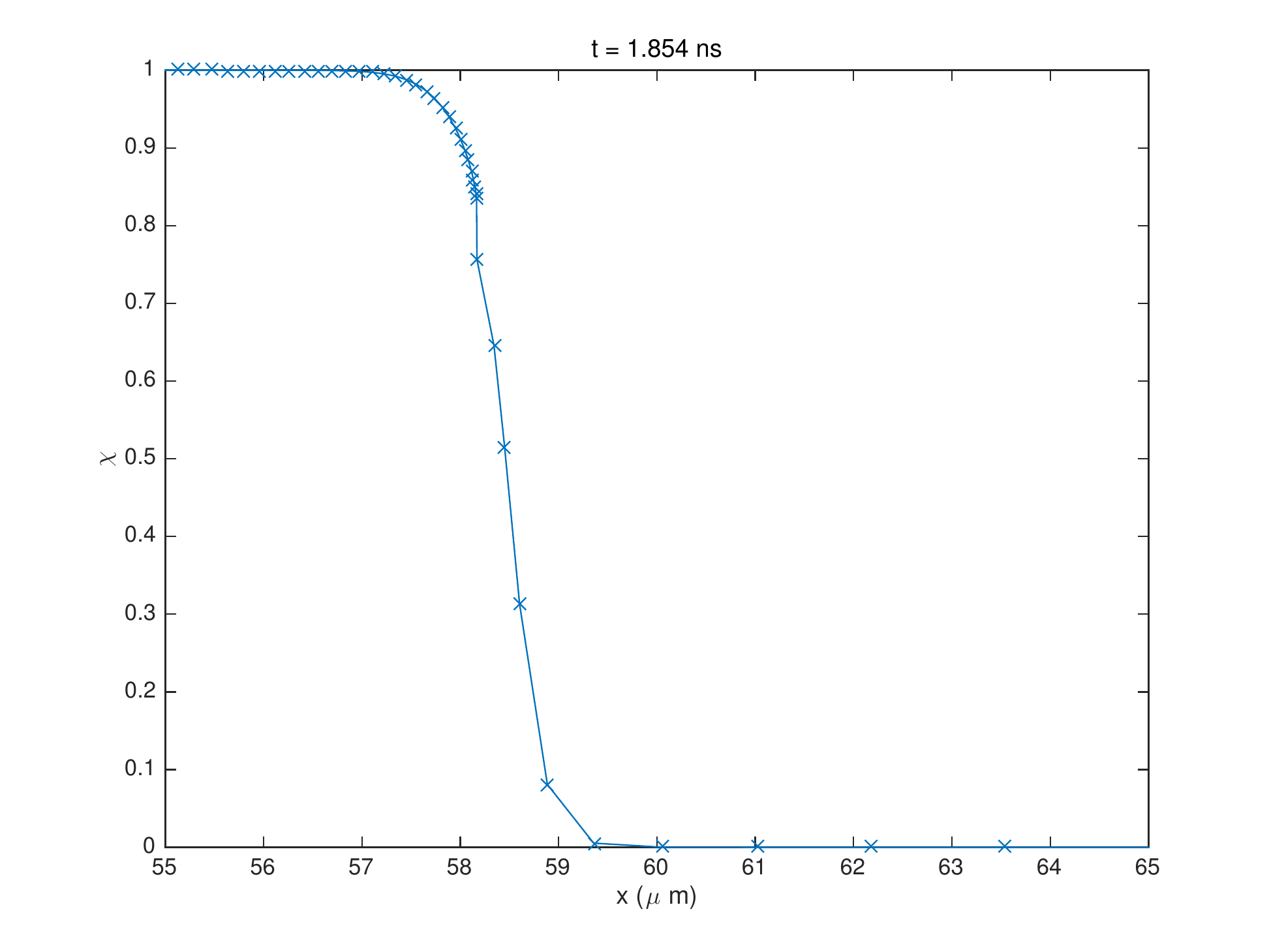}}
\caption{Plot of the mixing layer between DD and CH at the time of compression burn for (a) the Inviscid Simulation and (b) Viscous Simulation}
\label{chis}
\end{figure}

Figure \ref{chis} shows the mass composition profiles due to plasma species transport across the fuel-capsule  interface at the time of maximum compression, $t \approx 1.85 ns$. The radial locations of the midpoints  of the mixing layers at this time are 55 microns for the inviscid case, and 58.5 microns for the viscous case, where the difference of approximately 3.5$\mu m$ is 6.4\% of the inviscid radius and corresponds to a difference of 20\% in fuel volume. The mixing layer itself is slightly wider in the viscous case, $\approx 2.5 \mu m$, than in the inviscid case at $\approx 2 \mu m$.  With the differences in mix width and radius taken together this corresponds to a mixed region volume  about $40\%$ larger in the viscous case.   A 1D HELIOS simulation of similar ICF implosions \cite{vold-sherrill} showed  greater convergence ($r_{min} \approx 20 \mu m$), but post-processed calculations from those simulations showed good  agreement with the mass transport mix widths in this study, suggesting comparable density and temperatures near the fuel-capsule interface in the two cases.   The convergence differences can be related to laser input and initial fill pressure differences.  The widths  of these mix layers by plasma gradient driven transport is  small in these idealized 1D simulations, however, we expect the mass diffusivity will play an enhanced role in 2D and 3D \cite{weber2014inhibition, plasmaInRTnKH} where the fuel-capsule  interfacial area can be increased significantly by instability growth and possibly by turbulence.

We considered results in Ref.~\cite{molvigvolddoddwilks} as a guide for expected  asymmetry of the mass species mixing profiles.  Our profile results are more symmetric between the fuel and capsule than expected, suggesting that the carbon-based capsule may not be sufficiently high Z, to achieve the asymmetry predicted in \cite{molvigvolddoddwilks}.

\textit{Results Summary. }
Select results from the inviscid and viscous simulations to 2.5 ns are summarized in Table I.  The values  indicate for the inviscid case that both maximum temperatures and densities at the origin occur during the 'spike' at first shock convergence ($t \approx 1.43 ns$) leading to high pressures in that spike.  The viscous case maximum densities and temperatures at the origin occur during compression and at slightly different times,  1.83 ns and 2.0 ns respectively.  The offset in time between the maxima in density and temperature leads  to a greatly reduced instantaneous pressure maximum at the origin.  \\\

\begin{table}[H]
	\label{basetable}
	\caption{ICF Result Summary}	
    \begin{center}
	\begin{tabular} {    l    |   l       l        }

		\hline
Model Parameter   & Inviscid case   & Viscous case   \\ \hline
Final time -ns-                         &     2.50   &   2.50       \\ \hline
 Highest temperature (ion) at origin  &  10392.35    & 4177.48    \\ \hline
 Highest temperature (e) at origin  &   1689.5     & 1579.69    \\ \hline
 Highest temperature occurs at time -ns-      &   1.430  & 1.833   \\ \hline
 Highest density at origin $10^{23} cm^{-3}$          &   128.95  &   3.216   \\ \hline
 Highest density occurs at time -ns-        &   1.430  &  2.027    \\ \hline
  Maximum pressure at origin (Gbar)  &    117.31   &  1.4139   \\ \hline
Ion Temp in Zone 1 at final time (eV)             &   405.06      & 433.72     \\ \hline
 Elec Temp in Zone 1  at final time (eV)           &  415.67    &  446.37    \\ \hline
 Rad Temp in Zone 1  at final time (eV)             &   425.91    &  454.92   \\ \hline
 Ion Temp in Outer Fuel Zone  at final time (eV)    &   152.7    &  162.2    \\ \hline
 Elec Temp in Outer Fuel Zone  at final time (eV)  &    152.6    &  162.1   \\ \hline
 Rad Temp in Outer Fuel Zone  at final time (eV)    &    152.7     & 162.3     \\ \hline
 Burn Weighted Ion Temperature (eV) &   2704.  & 2655.   \\ \hline



	\end{tabular}
	\end{center}
\end{table}

At the final run time, 2.5 ns, the three temperatures at the origin are similar but not completely equilibrated in the inviscid or viscous case.  The temperatures are slightly greater in the viscous cases, as expected due to the viscous heating.   At this late time, in the outer fuel zone at the fuel-plastic boundary, the three  temperatures have equilibrated in the inviscid and viscous cases with a slightly higher final temperature for the viscid case.  The  time averaged burn weighted ion temperatures are nearly identical for the viscid and inviscid cases, in sharp contrast to the maximum temperature values which spiked at the origin for the inviscid case.

\textit{Artificial viscosity.}
Trade-offs between physical plasma viscosity and artificial viscosity in Lagrange numerical methods have been examined in \cite{mason2014real}.  Our results compliment and extend their findings.  
We found that with plasma viscosity in the simulation, the solution remains numerically stable if the artificial viscosity is turned off after the shock converges on the target center.  After this time,  the entire domain has become a fully ionized plasma with significant viscosity.  In the simulations including plasma viscosity  and with  artificial viscosity turned off, the solutions show considerably more fluctuations compared to the inviscid or the viscous case when artificial viscosity is included. These fluctuations are seen  in the fuel and capsule regions especially as the shocks pass  from the center back into the capsule material.  It is hypothesized that the increased fluctuations, apparent in the absence of artificial viscosity, lead to increases  in kinetic energy and entropy which can reduce implosion efficiency.   This result was observed near the end of the current project and is planned to be explored  in future work.

\section{Conclusion}

In this study, a one dimensional,  staggered grid Lagrangian hydrodynamic code was written to include three temperature effects for ions, electrons, and radiation with plasma transport and dissipation including  viscous diffusion of momentum, viscous heating of ions, and species mass transport, in addition to the usual plasma physics describing  species thermal conductivities and  temperature coupling through collisional equilibration.  This code was used to examine ICF implosions typical of certain  DD-CH fuel-capsules  at the Omega Laser Facility.

The goal was to show first that the code recovers reasonable and realistic results compared to experiments and other codes for this class of ICF implosions while then examining the differences in result details for A-B comparison simulations which differ only in the plasma viscosity.  In each of these cases, the incidence of the first shock on target center occurs at 1.4 - 1.5 ns in the simulations and the return shock arrives at 1.8 - 1.9 ns.  These shock timings are in generally good agreement with previous calculations and measurements  \cite{soures1996direct, radha2005multidimensional, HELIOS, vold-sherrill, welser1, welser2, dodd2012effects}.    The neutron production rate, and the maximum and burn weighted ion temperatures  match expected results  \cite{similarBurnandT, bradley, meyerhofer}.  The minimum radius of the fuel during compression reached $\approx 50 \mu m$, in contrast to  previous  calculations of  $ \approx 20 ~ \mu m$ using HELIOS \cite{welser1, welser2, vold-sherrill}, which can be related to differences in the prescribed laser inputs, thickness of the capsule, and fuel fill pressures.

Our examination of viscosity in ICF implosions, shows the plasma viscosity and viscous heating of the ions resulted in differences in maximum fuel compression, neutron production rate, and the time dependent burn weighted ion temperatures.   The viscous shock front is about $8 \%$ ahead of the inviscid shock  at the  early time (1 ns) position  of the incoming shocks.  While the inviscid shock is slower at 1 ns, it  arrives earlier on axis by approximately 0.1 ns.  This behavior may be related to the radial convergence terms in the viscous dissipation which play a minor role at the large radius at early time, but play a more important role as the convergence becomes significant   at the times of first shock collapse on target center and during maximum compression.

The fuel compresses to a minimum radius for the viscous case that is greater by 3.5  $\mu m$, or $6\%$, compared to the inviscid  compression.  This  corresponds to a $20\%$ increase in fuel volume for the viscous case, and combined with a slightly wider mix width for the viscous case there is a $40\%$ increase in the mix volume for the viscous case.  The widths for the mass composition mixing layer of a few microns calculated from post-processed HELIOS simulations \cite{vold-sherrill, welser1} are in general agreement with the viscous and the inviscid results here.  It is evident from the results that viscous effects cause a decrease in the compression.  This increase in the size of the compressed region with viscosity reduces the \textit{rho-r} of the fuel and is also related to an increase in entropy deposited in the fuel and capsule due to the viscous dissipation through both the momentum and energy equations.  It is well known that entropy production degrades confinement and is a key concern in implosion efficiency  \cite{goncharov}. 

The ion temperatures are consistently higher for the  inviscid case at first shock convergence.  This leads to a factor of approximately  two difference in the ion temperatures at that time and  it follows that the neutron production rate is also augmented during shock convergence.  The time averaged burn weighted ion temperatures are only slightly  larger  in the inviscid case and at maximum compression they agree within a few percent.  The neutron yields are comparable during this phase between inviscid and viscous cases.   
It was also determined that the implementation of viscous effects modifies the need for an electron heat flux limiter in these hydrodynamic simulations.   As the viscous terms  increase the width of the shock front and decrease the magnitude of the ion temperature in the shocked region, the need for a flux limiter may be reduced across the domain with viscosity included.

After the first shock converges, the entire region becomes a plasma with significant temperatures driving up the magnitude of the plasma viscosity.  After that time, the plasma viscosity and viscous dissipation eliminate the need for artificial viscosity to stabilize the numerics, and the artificial viscosity can be turned off.  Results suggest that the subsequent solutions allow  greater shock-wave fluctuations in the fuel and especially in the capsule material compared to cases when artificial viscosity is included.  Previous work in this area \cite{mason2014real} confirms this is a topic worth examining further and in greater detail.

It was determined that viscous effects play a crucial role in implosion dynamics for this DD-CH model suggesting that other codes should consistently include viscous effects to model ICF implosions with greater fidelity.  Viscosity influences shock timings, fuel compression, neutron production rates, and heat fluxes especially at the time of first shock convergence.  Viscous effects during the later compression phase appear to be small in this 1D example, but are likely to play a more important role in 2D and 3D as suggested in previous studies \cite{weber2014inhibition, plasmaInRTnKH}.  It is expected  that  viscosity and plasma species mass transport  will  have a significant influence in other ICF capsule implosions.


\section{Acknowledgments}
\label{acknowledgements}
Much of this work was performed in the Los Alamos Summer Students Computational Physics Workshops  during the summers of 2013 and 2014.  The workshops are directed by  Dr.~S.~Runnels, and the authors thank him for providing this opportunity.
Los Alamos National Laboratory is operated by Los Alamos National Security, LLC for the U.S.~Department
of Energy NNSA under Contract No. DE-AC52-06NA25396.





\begin{thebibliography}{10}


\bibitem{Atzenitext}
S.~Atzeni, and J. Meyer-ter-Vahn. 
\newblock The Physics of Inertial Fusion: Beam Plasma Interaction, Hydrodynamics, Hot Dense Matter.
\newblock {\em  Clarendon Press}, Gloucestershire, UK  (2004).


\bibitem{soures1996direct}
J.M. Soures,  R.L. McCrory,   C.P. Verdon,  A.  Babushkin,   R.E. Bahr,  T.R. Boehly,   et.al.,
\newblock  Direct-drive laser-fusion experiments with the OMEGA, 60-beam, 40 kJ, ultraviolet laser system.
\newblock {\em Phys.Plasmas} {\bf{21(2)}} 2108  (1996).


\bibitem{molvigvolddoddwilks}
K.~Molvig,  E.~L.~Vold, E.~S~Dodd, and S.~C.~Wilks,
\newblock Non-Linear Structure of the Diffusing Gas-Metal Interface in a Thermonuclear Plasma.
\newblock {\em Phys. Rev. Lett.} {\bf{113}} 145001 (2014).


\bibitem{amendt2011potential}
P.~Amendt, S.~C.~ Wilks, C.~Bellei, C.~K.~Li, and R.~D.~Petrasso, 
\newblock  The potential role of electric fields and plasma barodiffusion on the inertial confinement fusion.
\newblock {\em Phys. Plasmas} {\bf{18(5)}} 056308  (2011).


\bibitem{schurtz}
G.P. Schurtz, P. D. Nicolai, and M. Busquet,
\newblock A nonlocal electron conduction model for multidimensional radiation hydrodynamics codes.
\newblock  {\em Phys. Plasmas} {\bf{7(10)}}   4238 (2000).

\bibitem{schurtz2007revisiting}
G. Schurtz, S.Gary, S Hulin, C  Chenais-Popovics, J.C.~Gauthier, F Thais, et.al.,
\newblock Revisiting nonlocal electron-energy transport in inertial-fusion conditions.
\newblock {\em Phys. Rev. Lett.} {\bf{98(9)}} 095002  (2007).


\bibitem{marocchino2014effects}
A.S. Marocchino,  S. Atzeni,  and A. Schiavi,
\newblock  Effects of non-local electron transport in one-dimensional and two-dimensional simulations of shock-ignited inertial confinement fusion targets.
\newblock {\em Phys. Plasmas} {\bf{21(1)}} 012701  (2014).


\bibitem{radha2005multidimensional}
P.B. Radha,   T.J.B. Collins, J.A.  Delettrez,  Y. Elbaz, R. Epstein, V.Yu Glebov, V.N. Goncharov, R.L. Keck, J.P. Knauer, J.A.  Marozas,  et.al.,
\newblock Multidimensional analysis of direct-drive, plastic-shell implosions on OMEGA.
\newblock {\em Phys. Plasmas}  {\bf{12(5)}} 056307 (2005).


\bibitem{Theobald}
W.Theobald, R. Nora, M. Lafon, A. Casner, et.al.,
\newblock Spherical shock-ignition experiments with the 40 + 20 beam configuration on OMEGA.
\newblock {\em  Phys. Plasmas},  {\bf 19} 102706 (2014).

\bibitem{rinderknecht}
H.G.Rinderknecht, H.Sio, C.K.Li,N.Hoffman, et. al.
\newblock Kinetic mix mechanisms in shock-driven inertial  confinement fusion implosions.
\newblock {\em  Phys. Plasmas},  {\bf 21} 056311; doi 10.1063/1.4876615   (2014).


\bibitem{HoffmanEtAl}
N.~Hoffman, G.B.~Zimmerman, K.~Molvig, H.G.~Rinderknecht, M.J.Rosenberg, et. al., 
\newblock Approximate models for the ion-kinetic regime in ICF capsule implosions.
\newblock {\em Phys.~Plasmas}  {\bf{22}} 052707 (2015).

\bibitem{gasmetal}
K.~Molvig, A.~N.~Simakov, and E.~L.~Vold,
\newblock Classical transport equations for burning gas-metal plasmas.
\newblock {\em Phys.~Plasmas}  {\bf{21}} 092709 (2014).


\bibitem{Simakov}
A.~N.~Simakov, and K.~Molvig,
\newblock Electron transport in a collisional plasma with multiple ion species.
\newblock {\em Phys.~Plasmas} 21:024503, (2014).


\bibitem{hinton}
F.~L.~Hinton,
\newblock Collisional Transport in Plasma, p.147 in,  {\em Handbook of Plasma Physics, Vol.I}, 
\newblock ed. A.A.~Galeev, R.N.~Sudan, North Holland Pub., New York, (1983).

\bibitem{spitzer}
L.~Spitzer, Jr.,
\newblock Physics of Fully Ionized Gases.
\newblock {\em  Dover edition,}  revised 1962 2nd ed., Interstellar Sci., NY (2006).

\bibitem{Atzeni87}
S.~Atzeni,
\newblock The physical basis for numerical fluid simulations in laser fusion.
\newblock {\em  Plasma Phys.~Control.~Fusion},  {\bf 29}(11):1535-604, (1987).

\bibitem{wilkins} 
M.L..~Wilkins,
\newblock Use of artificial viscosity in multidimensional fluid dynamic calculations.
\newblock {\em J. Comput. Phys.}  {\bf{36}} 281-303 (1980).




\bibitem{ThomasKares}
V.~Thomas, and R.~Kares,
\newblock Three-Dimensional Simulation Strategy to Determine the Effects of Turbulent Mixing on Inertial-Confinement-Fusion Capsule Performance.
\newblock {\em Phys. Rev. Lett}, {\bf 109}:075004, (2012).


\bibitem{hainesicf}
B.~M.~Haines, F.~F.~Grinstein, and J.~R.~Fincke,
\newblock Three-Dimensional Simulation Strategy to Determine the Effects of Turbulent Mixing on Inertial-Confinement-Fusion Capsule Performance.
\newblock {\em Phys. Rev. E}, {\bf 89}:053302, (2014).

\bibitem{WilsonEtAl}
D.C.~Wilson, P.S.~Ebey, T.C.~Sangster, W.T.~Shmayda, and V.Yu.~Glebov, 
\newblock Atomic mix in directly driven inertial confinement implosions.
\newblock {\em  Phys. Plasmas},  {\bf 18} 112707,  (2011).


\bibitem{Yabe}
T.~Yabe, and K.A.~Tanaka,
\newblock Long ion mean free path and nonequilibrium radiation effects on high-aspect-ratio laser-driven implosions.
\newblock {\em  Laser and Particle Beams},  {\bf (7)2} p.259-265 (1989).

\bibitem{Vidal}
F.~Vidal, M.Matte, M. Casanova, and O. Larroche,
\newblock Ion kinetic simulations of the formation and propagation of a planar collisional shock wave in a plasma.
\newblock {\em  Phys. Fluids B},  {\bf (5)9} 3182 (1993).



\bibitem{manheimer2007effects}
W.~Manheimer, and D.~Colombant,
\newblock Effects of viscosity in modeling laser fusion implosions.
\newblock {\em Laser and Particle Beams}, {\bf{25(04)}}, 541-547, (2007).


\bibitem{vold-sherrill}
E.~Vold, and L.~Welser-Sherrill,
\newblock Momentum Transport and Associated Scale Lengths in an ICF Plasma.  Presentation to APS DPP, Salt Lake City, Nov., 2011,  {\em Los Alamos Laboratory Report LA-UR-11-06397},  
\newblock and (Full Report), {\em Los Alamos Laboratory Report LA-UR-11-06396}, Los Alamos, NM October, (2010).


\bibitem{weber2014inhibition}
C.~R.~Weber, D.~S.~Clark, A.~W.~Cook, L.~E.~Busby, and H.~F.~Robey.
\newblock Inhibition of turbulence in inertial-confinement-fusion hot spots by viscous dissipation.
\newblock {\em Phys.~Rev.~E}, {\bf 89}:053106, (2014).


	
\bibitem{plasmaInRTnKH}
B.H.~Haines, E.L.~ Vold,  K.~Molvig, R.~Rauenzahn,  and C.~Aldrich.
\newblock Plasma Transport in Rayleigh-Taylor and Kelvin-Helmholtz Instabilities.
\newblock {\em Phys. Plasmas}, {\bf{21(9)}}, 092306,  DOI: 10.1063/1.4895502   (2014).



\bibitem{mason2014real}
R.J.~Mason, R.C.~Kirkpatrick, and R.J.~Faehl,
\newblock  Real viscosity effects in inertial confinement fusion target deuterium--tritium micro-implosions.
\newblock {\em Phys. Plasmas} {\bf{21}} 022705  (2014), and  {\em Phys. Plasmas} {\bf{21}} 039902  (2014).



\bibitem{Braginskii}
S.~I.~Braginskii.
\newblock Transport Processes in a Plasma, p.205
\newblock in {\em Reviews of Plasma Physics, Vol.~I}, edited by M.~A.~Leontovich, Consultants Bureau, New York, (1965).


\bibitem{nrl}
J.~D.~Huba,
\newblock NRL Plasma Formulary.
\newblock {\em http://wwwppd.nrl.navy.mil/nrlformulary/}, US Navel Research Lab, Washington, D.C., revised (2013).


\bibitem{KaganTang1}
G.~Kagan, and X-Z.~Tang,
\newblock Electrodiffusion in a plasma with two ion species.
\newblock {\em Phys.~Plasmas}  {\bf{19}} 082709 (2012).

\bibitem{KaganTang2}
\newblock G.~Kagan, and X-Z.~Tang,
\newblock Thermodiffusion in inertially confined plasma.
\newblock {\em Phys.~Lett.~A}  {\bf{378}} 1531-35 (2014).


\bibitem{vold2010dtburn}
E.~Vold, J.~Hansen, N.~ Hryniw, L.A.~ Kesler, and F.~ Li,
\newblock Uniform DT 3T Burn: Computations and Sensitivities.
\newblock {\em Los Alamos Laboratory Report LA-UR-11-00655} proceedings of the NECDC Conference, Los Alamos, NM October, (2010).



\bibitem{MolvigAlme}
K.~Molvig, M.~Alme, R.~Webster, and C.~Galloway,
\newblock Photon coupling theory for plasmas with strong Comptonscattering: four temperature theory.
\newblock {\em Phys. Plasmas}  {\bf{16}}  023301 (2009).


\bibitem{FennMollVold}
D.~Fenn, R.~Moll, and E.~Vold,
\newblock Plasma Mixing in ICF Applications.
\newblock {\em Los Alamos Laboratory Report LA-UR-13-26576}, Los Alamos Computational Physics Workshop 2013, Los Alamos, NM October, (2010).



\bibitem{Macfarlane}
J.J~MacFarlane, I.E.~Golovkin, and P.R.Woodruff,
\newblock HELIOS-CR: A 1-D radiation-magnetohydrodynamics code with inline atomic kinetics modeling.
\newblock {\em J. Quant. Spectroscopy \&Rad. Transfer}  {\bf{99}} 381-397 (2005).


\bibitem{goncharov}
V.N.~Goncharov, O.V.~Gotchev, E.~Vianello, et.al.,
\newblock Early stage of implosion in inertial confinement fusion: Shock timing and perturbation evolution.
\newblock  {\em Phys. Plasmas} {\bf{13}}   012702 (2006).


\bibitem{trubnikov}
B.~A.~Trubnikov,
\newblock Particle Interactions in a Fully Ionized Plasma.~p.105 in,
\newblock {\em Reviews of Plasma Physics, Vol.~I}, edited by M.~A.~Leontovich, Consultants Bureau, New York, (1965).






	
\bibitem{dolan1982fusion}
T.J.~Dolan, 
\newblock {\em Fusion Research}, Pergamon Publishing, New York, (1982).





\bibitem{welser1}
L. Welser-Sherrill, J. H. Cooley, D. A. Haynes, D. C. Wilson, M. E. Sherrill, R. C. Mancini, and R. Tommasini,
\newblock  Application of fall-line mix models to understand degraded yield.
\newblock   {\em  Phys. Plasmas}   {\bf{15}}, 072702  (2008).


\bibitem{welser2}
L. Welser-Sherrill, D.A. Haynes, R.C. Mancini, J.H. Cooley, R. Tommasini, I.E. Golovkin,
M.E. Sherrill, and S.W. Haan,
\newblock  Inference of ICF implosion core mix using experimental data and theoretical mix modeling.
\newblock   {\em  High Energy Density Physics}   {\bf{5}},  249Ð257  (2009).


\bibitem{dodd2012effects}
E.S.~Dodd, J.F.~Benage,  G.A.~Kyrala, D.C.~Wilson, F.J.~Wysocki,  W.~Seka,  V.Yu Glebov,  C.~Stoeckl and J.A.~Frenje,
\newblock The effects of laser absorption on direct-drive capsule experiments at OMEGA.
\newblock {\em Phys Plasmas} {\bf{9(4)}} 042703 (2012).



\bibitem{betti2007shock,}
R.~Betti, C.D.~Zhou,  K.S.~Anderson,  L.J.~Perkins,  W.~Theobald,  and A.A.~Solodov, 
\newblock Shock ignition of thermonuclear fuel with high areal density.
\newblock {\em Phys. Rev. Lett.} {\bf{98(15)}} 155001,   (2007).


\bibitem{HELIOS}
I.E.~Golovkin,  J.J.~MacFarlane, P.R.~Woodruff, L.A.~Welser,  D.L.~McCrorey,  R.C.~Mancini,  and J.A.~Koch, 
\newblock Modeling of indirect-drive ICF implosions using 1D hydrodynamic code with inline collisional-radiative atomic kinetics. 
\newblock {\em Inertial Fusion Science and Applications 2003} 166-169  (2004).

\bibitem{similarBurnandT}
D. C. Wilson, C. W. Cranfill, C. Christensen, R. A. Forster, R. R. Peterson, N. M. Hoffman, and G. D. Pollak,
C. K. Li, F. H. Seguin, J. A. Frenje, and R. D. Petrasso, 
P. W. McKenty, F. J. Marshall, V. Yu. Glebov, and C. Stoeckl,
G. J. Schmid, N. Izumi, and P. Amendt, 
\newblock Multifluid interpenetration mixing in directly driven inertial confinement fusion capsule implosions.
\newblock   {\em Phys. Plasmas} {\bf{11(5)}}   2723 (2004).


\bibitem{bradley}
P.~Bradley, 
The effect of mix on capsule yields as a function of shell thickness and gas fill.
\newblock   {\em Phys. Plasmas} {\bf{21}}   062703; doi: 10.1063/1.4882247 (2014).

\bibitem{meyerhofer}
D. D.~Meyerhofer, J.A.~Delettrez, R.~Epstein, V.Y.~Glebov., V.N.~Goncharov, R.L.~Keck, É and T.C.~Sangster, 
Core performance and mix in direct-drive spherical implosions with high uniformity.  {\em Phys. Plasmas},  {\bf{8(5)}}, 2251  (2001). 

\bibitem{Li}
C.K.~Li, F.H.~Seguin, J.~Frenje, S.~Kurebayashi, R.D.~Petrasso, D.D.~Meyerhofer, Éand C.~Stoeckl,  Effects of fuel-shell mix upon direct-drive, spherical implosions on OMEGA.  {\em Phys. Rev. Lett.},  {\bf{89(16)}}, 165002 (2002). 





\end{thebibliography}

\end{document}